\documentclass{aa}  

\usepackage{graphicx}
\usepackage{float}
\usepackage{sidecap} 
\usepackage{txfonts}
\usepackage{color}

\usepackage{enumitem}
\usepackage{ulem}

\usepackage{pifont} 

\usepackage{color}

\usepackage{tikz}

\newcommand\kms{km\,s$^{-1}$}

\newcommand\Mw{\dot{M}_{\rm w}}
\newcommand\Mco{\dot{M}_{\rm obs}}
\newcommand\Msun{$M_{\odot}$}

\newcommand\Hy{\text{H}}

\newcommand\dens{$~\rm{cm}^{-3}$}
\newcommand\nH{n_{\text{H}}}

\newcommand{\BG}[1]{\textcolor{black}{#1}} 

\newcommand{\SC}[1]{\textcolor{black}{#1}}
\newcommand{\REV}[1]{\textcolor{black}{#1}}

\begin{document}

   \title{Molecule formation in dust-poor irradiated jets}
 \subtitle{I. Stationary disk winds}
   \author{B. Tabone 
          \inst{1, 2} \thanks{tabone@strw.leidenuniv.fr}, B. Godard \inst{2, 3}, G. Pineau des Forêts \inst{2, 4}, S. Cabrit\inst{2}, E. F. van Dishoeck \inst{1, 5} 
          }
   \institute{Leiden Observatory, Leiden University, PO Box 9513, 2300 RA Leiden, The Netherlands
                       \and
                       LERMA, Observatoire de Paris, PSL Research Univ., CNRS, Sorbonne Univ., 75014 Paris, France
                       \and
                       Laboratoire de Physique de l’\'{E}cole normale supérieure, ENS, Université PSL, CNRS, Sorbonne Universit\'{e}, Universit\'{e} Paris-Diderot, Sorbonne Paris Cit\'{e}, 75005 Paris, France
                       \and 
             Université Paris-Saclay, CNRS, Institut d’Astrophysique Spatiale, 91405, Orsay, France
             \and
             Max-Planck-Institut für Extraterrestrische Physik, Giessenbachstrasse1, 85748 Garching, Germany
        }   
\date{\today}
\abstract {Recent ALMA observations suggest that the highest velocity part of molecular protostellar jets ($\gtrsim 80$~km s$^{-1}$) are launched from the dust-sublimation regions of the accretion disks ($\lesssim 0.3$ au). However, formation and survival of molecules in inner protostellar disk winds, in the presence of a harsh far-ultraviolet (FUV) radiation field and the absence of dust, remain unexplored.
}
{We aim at determining if simple molecules such as H$_2$, CO, SiO, and H$_2$O can be synthesized and spared in fast and collimated dust-free disk winds or if a fraction of dust is necessary to explain the observed molecular abundances.}
{This work is based on a recent version of the Paris-Durham shock code designed to model irradiated environments. Fundamental properties of the dust-free chemistry are investigated from single point models. A laminar 1D disk wind model is then built using a parametric flow geometry. This model includes time-dependent chemistry and the attenuation of the radiation field by gas-phase photoprocesses. The influence of the mass-loss rate of the wind and of the fraction of dust on the synthesis of the molecules and on the attenuation of the radiation field is studied in detail.}
{We show that a small fraction of H$_2$ ($\le 10^{-2}$), primarily formed through the H$^-$ route, can efficiently initiate molecule synthesis such as CO and SiO above $T_{\rm{K}}\sim$ 800~K. We also propose new gas-phase formation routes of H$_2$ that can operate in strong visible radiation fields, involving for instance CH$^+$. The attenuation of the radiation field by atomic species (eg. C, Si, S) proceeds through continuum self-shielding. This process ensures efficient formation of CO, OH, SiO, H$_2$O through neutral-neutral reactions, and the survival of these molecules. Class 0 dust-free winds with high mass-loss rates ($\dot{M}_{\text{w}} \ge 2 \times 10^{-6}$ $M_{\odot}~\rm{yr}^{-1}$) are predicted to be rich in molecules if warm ($T_{\rm{K}}\ge 800$~K). Interestingly, we also predict a steep decrease in the SiO-to-CO abundance ratio with the decline of mass-loss rate, from Class 0 to Class I protostars. The molecular content of disk winds is very sensitive to the presence of dust and a mass-fraction of surviving dust as small as $10^{-5}$ significantly increases the H$_2$O and SiO abundances.}{Chemistry of high velocity jets is a powerful tool to probe their content in dust and uncover their launching point. Models of internal shocks are required to fully exploit the current (sub-)millimeter observations and prepare future JWST observations.}


\maketitle
   

\section{Introduction}

Protoplanetary disks provide their host accreting star with material and regulate the formation, growth, and migration of planets. The global evolution and dispersal of disks around nascent stars is regulated by the transport of angular momentum and mass-loss processes \citep{2011ARA&A..49..195A}. Fast jets\footnote{In this paper, "jet" refers to observed fast ($\gtrsim 50$~km~s$^{-1}$) and collimated (opening angle $\lesssim 8^{\circ}$) outflowing gas whereas "wind" refers to theoretical models that account for the origin of jets.} are ubiquitously observed in accreting young stars of all ages, with universal collimation and connection between accretion and ejection, probably of magnetic origin \citep{2002EAS.....3..147C,2007IAUS..243..203C}. However, the region of the disk actually involved in mass ejection,  
and the associated angular momentum extraction, remain a topic of hot debate.

The main observational method proposed so far to locate the launching region of jets relies on the joint measurement of rotation and axial velocities \citep{2003ApJ...590L.107A,2006A&A...453..785F}. While studies of atomic jets are currently limited by spectral resolution in the optical range \citep{2016ApJ...832..152D}, the unique combination of high spectral ($<0.5$\kms) and spatial ($50~$mas) resolution \SC{of ALMA} now allows to conduct similar tests on jets from the youngest protostars, so-called Class 0, which are much brighter in molecules \citep[eg.][]{2010A&A...522A..91T} than jets from more evolved protostars and pre-main sequence stars (Class I and II) which are mainly atomic. 
In this context, high angular resolution observations ($\simeq 8~$au) of the fastest part of the HH~212 jet have unveiled rotation signatures in SiO emission suggestive of a disk wind launched within $0.3$~au \citep{2017NatAs...1E.152L,2017A&A...607L...6T}. Owing to the high bolometric luminosity of the central protostar \citep[$L_{bol} \simeq 9 L_{\odot}$, ][]{1992A&A...265..726Z}, dust is expected to be sublimated \SC{within 0.3} au, and the HH212 SiO-rich jet would thus trace a dust-free magnetohydrodynamic (MHD) disk wind.

\SC{To test the likelihood of this interpretation, it is now} of paramount importance to check if the presence of \SC{SiO} molecules in the jet of HH~212 is indeed compatible with a dust-free disk wind origin, as suggested by the \SC{rotation} kinematics.
Detailed astrochemical modeling of dusty magnetized disk winds show that molecules can survive the acceleration and the far-ultraviolet (FUV) field emitted by the accreting protostars \citep{2012A&A...538A...2P}. However, molecule formation in dust-free disk winds remains an open question. In the absence of dust, the FUV field can \SC{more easily} penetrate the unscreened flow and photodissociate molecules, whereas H$_2$ formation on grains, the starting point of molecule synthesis, is severely \SC{reduced}. So far, astrochemical models  \SC{have only investigated dust-free} winds launched from the stellar surface.
Pioneering studies have shown that \SC{they} are hostile to the formation of H$_2$, due to photodetachment of the key intermediate H$^-$ by visible photons from the star \citep{1988MNRAS.230..695R,1989ApJ...336L..29G,1990ApJ...361..546R}, whereas other molecules such as CO, SiO or H$_2$O are destroyed when a \SC{strong}
UV excess is included \citep{1991ApJ...373..254G}.
\SC{However, there have been yet no similar investigations in a disk wind geometry, and with a fully self-consistent FUV field.}

The origin of the observed molecules in Class 0 jets is \SC{actually} still debated, as high-velocity molecular emission is not necessarily tracing a pristine wind. Instead of assuming that molecules are material ejected from the vicinity of the star ("wind" scenario), another class of scenarios proposes that molecules are "entrained" by a fast and unseen atomic jet through bow-shocks or turbulence \citep{1993A&A...278..267R,1991ApJ...372..646C}. Even if observations of small-scale molecular micro-jets \citep[$\simeq 10-400~$au,][]{2007A&A...468L..29C,2017NatAs...1E.152L} may contradict entrainment of envelope material, a slow dusty disk wind surrounding the fast jet could still bring \SC{fresh} molecular material close to the jet axis and \SC{explain} the observed collimation properties \SC{of molecular jets} \citep{2016MNRAS.455.2042W,2018A&A...614A.119T}. Observational diagnostics \SC{of the wind launch radius} based on \SC{rotation signatures would not be reliable anymore} in such time-variable jets \citep[eg.][]{2011ApJ...737...43F} or in jets prone to turbulent mixing.

 In contrast with an inner disk wind scenario, the entrainment scenario implies that molecular material is rich in dust. Chemistry can then be used as a powerful diagnostic. Early astrochemical models of stellar winds already pointed-out unique features of dust-free chemistry such as a low CO/H$_2$ ratio \citep{1989ApJ...336L..29G}. It suggests that chemistry is a promising diagnostic of the dust content, and as such, could distinguish between dust-free disk wind and entertainment scenario. However, because sublimation temperature depends on the composition and sizes of grains, jets launched from the dust sublimation region of silicate and carbonaceous grains may contain a small fraction of surviving dust such as aluminum oxide grains (Al$_2$O$_3$) for which sublimation temperature is higher, $\simeq 1700~$K \citep{1995ApJ...447..848L}. Despite representing a small mass-fraction of the total interstellar dust \citep[$\simeq 2\%$, assuming solar elemental abundances,][]{2009ARA&A..47..481A}, aluminum oxide grains could have a strong impact on the chemistry and bias the proposed test. Understanding the precise impact of a non-vanishing fraction of dust on the chemistry is thus an important step to distinguish "wind" to "entertainment" scenario.

Determining if molecular jets are indeed launched within the dust sublimation radius is a crucial question, as they would then bring new clues to planet formation theories. Recent studies suggest that the first steps of planet formation may occur in the protostellar phase \citep[eg.][]{2010MNRAS.407.1981G,2018A&A...618L...3M,2018NatAs...2..646H}.
Probing the bulk elemental composition and inner depletion pattern of the inner gaseous regions of protoplanetary disks within the dust sublimation radius is a powerful tool to uncover key disk processes related to planet formation such as dust trapping in the outer parts of the disks \citep{2019arXiv191007345M, 2019arXiv191006029M}. However, deeply embedded Class 0 disks are too extincted to be probed by optical and near-infrared atomic lines. Hence, if protostellar molecular jets are tracing a pristine dust-free disk wind, they would offer a unique opportunity to have access to the elemental composition of the inner region of Class 0 disks, and thus reveal elusive disk processes. In this perspective, observations of high velocity molecular bullets toward active protostars show abundant oxygen-bearing species (SiO, SO) but a drop in carbon-bearing species such as HCN or CS that was interpreted as a low C/O ratio \citep{2010A&A...522A..91T,2019arXiv191007857T}. However, it remains unclear if abundance ratios of molecular tracers are indeed directly related to a change in elemental abundances rather due to unique features of dust-poor chemistry or shocked gas. A fine understanding of the chemistry operating in dust-free or dust-poor jets is required to use molecular jets as a probe of elemental abundances of the inner disks.

In the present paper, we revisit pioneering astrochemical models of stellar winds by investigating if molecules can be formed in a disk wind launched within the dust sublimation radius. We focus our analysis on H$_2$ and on the most abundant oxygen-bearing molecules observed toward protostellar jets, namely CO, SiO, OH, and H$_2$O \citep[eg.][]{2010A&A...522A..91T,2011A&A...531L...1K}. 
Sulfur and nitrogen chemistry, as well as the dependency of molecular abundances on elemental ratios is beyond the scope of the present paper. Throughout this work, special attention is paid to the impact of a small fraction of dust on the chemistry to determine if molecular abundances can be used as a discriminant diagnostic even when the wind contains surviving refractory dust. The thermal balance, together with shock models will be presented in the next paper of this series. 

In Section \ref{sec:intro-model}, we present the basic physical and chemical ingredients of the astrochemical model. For sake of generality, we then study molecule formation with the use of single point models (Section \ref{sec:singlepoint}). It allows us to derive simple criteria for molecules to be abundant, as well as to specify their formation routes. More realistic models assuming a specific flow geometry are explored in Section \ref{sec:wind}. In contrast with single-point models, they include the effect of the dilution of the radiation field and density, time-dependent chemistry, and shielding of the radiation field. 
Limitations of the model, as well as observational diagnostics of dust-free and dust-poor winds are discussed in Section \ref{sec:disscussion}. Our findings are summarized Section \ref{sec:conclusion}.

\section{Numerical method}
\label{sec:intro-model}
Models presented throughout this work are based on the publicly available Paris-Durham shock code initially designed to compute the dynamical, thermal and chemical evolution of a plane-parallel steady-state shock wave. The code includes detailed microphysical processes and a comprehensive time-dependent chemistry \citep{2003MNRAS.343..390F,2013A&A...550A.106L}. The versatility of this code allows computing also the thermal and chemical evolution of any 1D stationary flow in a slab approach. In this work, the recent version developed by \citet{2019A&A...622A.100G}, which includes key processes of photon-dominated region (PDR) physics, has been further upgraded to ensure proper treatment of dust-free chemistry\footnote{\REV{the updated version of the code and former versions are available on the ISM platform \url{http://ism.obspm.fr}.}}. Two types of model have been computed: single spatial point models (Section \ref{sec:singlepoint}), and specific wind models (Section \ref{sec:wind}). We present here the basic numerical method used in this work. Details on the specific wind model, including the prescribed density structure, are given in Section \ref{subsec:themodel}.



\subsection{Geometry and parameters}

\begin{figure}
\centering
\includegraphics[width=.48\textwidth]{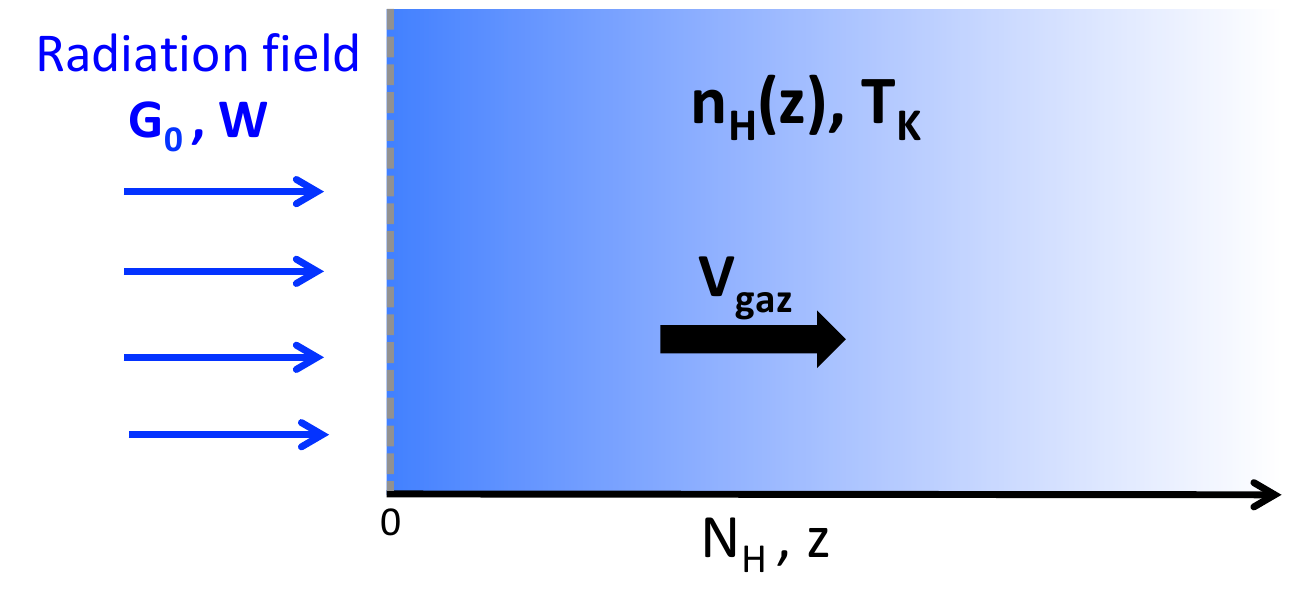} 
\caption{Schematic view of the 1D geometry used in this work. The code solves the chemical evolution of a slab of gas flowing at a constant velocity $V_{\text{gas}}$ and irradiated from the left (upstream) along a direction parallel to the flow. The proton density profile $\nH(z)$ is prescribed and the temperature $T_{\rm{K}}$ is constant. The radiation field is the sum of a FUV part modeled by an ISRF scaled by $G_0$ and a visible part modeled by a black body radiation field at $T_{\rm vis}=4000$~K diluted by a factor $W$ (see Appendix \ref{app:RF}). The attenuation of the radiation field by gas-phase photoprocesses and by dust (if any) along $z$ is computed consistently.}
\label{fig:shematic-view}
\end{figure}

Figure \ref{fig:shematic-view} gives a schematic view of the adopted geometry from which specific models can be built. The model is a slab of gas flowing at a velocity $V_{\text{gas}}$ along the direction $z$ and irradiated upstream ($z=0$ in Fig.\ref{fig:shematic-view}). The proton density $\nH(z) = n(\text{H})+2 n(\text{H}_2)+n(\rm{H}^+)$ along the slab is also prescribed. \REV{Throughout this work, the gas is assumed to be isothermal with the kinetic temperature $T_{\rm{K}}$}.
This allows us to study the influence of the temperature on the chemistry in a parametric way and independently of uncertainties in the thermal balance.

The impinging radiation field is parametrized as a sum of a FUV component modeled as a standard interstellar radiation field  \citep[ISRF, ][]{1983A&A...128..212M} scaled by a parameter $G_0$ plus a visible component modeled as a black-body radiation field at $T_{\rm vis}=4000$~K diluted by a factor\footnote{The shape of the visible part gives a reasonable approximation of the standard interstellar radiation field for $W \simeq 5 \times 10^{-13}$ (see Appendix \ref{app:RF}), and a good proxy for the visible field close to nascent low-mass stars with $W = 5 \times 10^{-7} (R/10~au)^{-2}$,
where R is the distance to the star and where we assume a stellar radius of $3 R_{\odot}$.} W. The adopted functional form of the FUV field, as well as the shape of the resulting unshielded radiation field,  are given in Appendix \ref{app:RF}.

\subsection{Radiative transfer and photodestruction processes}
\label{subsec:RT}

The attenuation of the radiation field between $\lambda = 91$ nm and $1.5\mu$m thought the slab is computed by considering absorption by continuous photoprocesses along rays perpendicular to the slab and parallel to the flow. The attenuation coefficient due to gas-phase processes is then
\begin{equation}
\kappa_G (z,\lambda) = \sum_{i} \sigma_i (\lambda) n_i (z),
\end{equation}
where $\sigma_i(\lambda)$ is the cross-section of the photoprocess involving the species X$_i$ and $n_i$ its particle number \REV{density}. When dust is included, \REV{we assume that the grain size follows a power-law distribution of index $-3.5$ with a minimum and maximum size of 0.01~$\mu$m and 0.3~$\mu$m, respectively \citep{1977ApJ...217..425M}}.
\REV{Following \citet{2019A&A...622A.100G} (see their Appendix B) and for simplicity,} absorption of UV photons by grains is calculated assuming the absorption coefficient of single size spherical graphite grains of radius $a_g = 0.02$~$\mu$m derived by  \citet{1984ApJ...285...89D} and \citet{1993ApJ...402..441L}, where $a_g \equiv \sqrt{<r_c^2>}$ \REV{is calculated from the mean square radius of the grains.}

Photodissociation and photoionization rates are computed following two different approaches. If the photoprocess leads to a significant attenuation of the FUV field or if it is a key destruction or formation pathway for a major species, then its rate is consistently computed by integrating the cross-section over the local radiation field including its UV and visible components (see Appendix \ref{app:RF}). 
Accordingly, the model includes absorption of photons by photoionization of C, S, Si, Mg, Fe, H$^-$, H$_2$O, SO, O$_2$, and CH, and photodissociation of OH, H$_2$O, SiO, CN, HCN, H$_2$O, H$_2^+$, SO, SO$_2$, O$_2$, CH$^+$ and CH. Cross-sections are taken from \citet{2017A&A...602A.105H} and subsequently resampled on a coarser grid \REV{of irregular sampling} (about $100$ points) to reduce computing time.

Alternatively, rates associated with other continuous photoprocesses are assumed to depend linearly on the integrated FUV radiation field as
\begin{equation}
k_{photo}= \alpha \frac{F(z)}{F_{\rm{ISRF}}},
\end{equation}
where $F_{\text{ISRF}}$ is the FUV photon flux from 911 \AA\ to 2400 \AA\ associated with the isotropic standard interstellar radiation field (ISRF), $F(z)$ is the local FUV photon flux computed over the same range of wavelength and $\alpha$ is the photodissociation rate for an unshielded ISRF. Note that even if this method seems to be crude, the non-trivial FUV attenuation by the dust-free gas prevents us from relying on more sophisticated fits as a function of $N_\text{H}$ as used in typical dusty PDR models.  

In order to avoid a prohibitive fine sampling of the radiation field around each UV line, CO and H$_2$ photodissociation are not treated according to the latter method. For CO photodissociation, we include self-shielding, and shielding by H$_2$ and by continuous process by expressing the photodissociation rate $k_{\rm{CO}}$ as
\begin{equation}
k_{\rm{CO}} = 2.06 \times 10^{-10} \text{s}^{-1} \theta_1(N(\text{CO})) \theta_2(N(\text{H}_2)) \frac{\chi}{1.23},
\end{equation}
where $\theta_1(N(\rm{CO}))$ and $\theta_2(N(\rm{H_2}))$ are self-shielding and cross-shielding factors tabulated by \citet{1996A&A...311..690L} and $\chi$ is the ratio at 100 nm of the local FUV flux to the mean interstellar radiation field of  \citet{1978ApJS...36..595D} ($2~\times 10^{-6}~\rm{erg~s^{-1}~\rm{cm}^{-2}}~\AA^{-1}$). The adopted shielding function gives a good approximation of the photodissociation rate of CO though neglecting the effect of the excitation of CO \citep{2009A&A...503..323V}. Note that the factor $1.23$ stands for the ratio between \citet{1983A&A...128..212M} and \citet{1978ApJS...36..595D} UV flux at 100~nm. The photodissociation rate of H$_2$ is computed under the FGK approximation \citep{1979ApJ...227..466F}.

\subsection{The excitation of H$_2$}

Time-dependent populations of the first 50 ro-vibrational levels of H$_2$ in the electronic ground state are computed, including
de-excitation by collision with H, He, H$_2$ and H$^+$ \citep{2003MNRAS.341...70F}, UV radiative pumping of electronic lines followed by fluorescence \citep{2019A&A...622A.100G}, and excitation of H$_2$  due to formation on grain surfaces. For H$_2$ formation on grain surface, we assume that the levels are populated following a Boltzmann distribution at a temperature of 1/3 of the dissociation energy of H$_2$ \citep{1987ApJ...322..412B}. For H$_2$ gas-phase formation routes, levels are populated in proportion to their local population densities.

\subsection{Elemental abundances}

\begin{table}[t]
\centering
\caption{Elemental abundances adopted in dust-free and dusty models. The last two columns give the distribution of elemental abundances for a standard dust-to-gas ratio $Q_{\text{ref}} \equiv 6 \times 10^{-3}$. For dusty models with different dust-to-gas ratio, total elemental abundances are kept constant and elemental abundances in the grains are reduced by the same factor. Total elemental abundances and the elemental abundance in the grains are taken from \citet{2003MNRAS.343..390F} where carbon and hydrogen locked in PAH are assumed to be released in gas-phase for all model. Numbers in parentheses are powers of 10.} 
 \begin{tabular}{c | c | c c } 
 \hline
   & $Q=0$ (dust-free) & $Q = Q_{\text{ref}} \equiv 6 \times 10^{-3}$ & \\ [0.5ex] 
 \hline
  Element  & Gas  &   Gas   & Grain \\
     \hline
 H  & 1.00 & 1.00 & 0.00 \\
 He & 1.00(-1) & 1.00(-1) & 0.00  \\
 C  & 3.55(-4) & 1.92(-4) & 1.63(-4)\\ 
 N  & 7.94(-5) & 7.94(-5) & 0.00 \\
 O  & 4.42(-4) & 3.02(-4) & 1.40(-4) \\
 Mg & 3.70(-5) & 0.0 & 3.70(-5) \\
 Si & 3.67(-5) & 3.03(-6) & 3.37(-5) \\
 S  & 1.86(-5) & 1.86(-5) & 0.00 \\
 Fe & 3.23(-5) & 1.50(-8)  & 3.23(-5) \\
  \hline\hline
\end{tabular}
\label{table:elementalabound}
\end{table}

Total fractional elemental abundances are constructed from Table 1 of \citet{2003MNRAS.343..390F}. For dust-free models, all elements are placed in the gas phase. For models with non-vanishing dust fraction, the dust content is quantified by the dust-to-gas mass ratio $Q$. The relative abundances between elements locked in the grains are assumed to be constant for any dust-to-gas ratio and equal to those of \citet{2003MNRAS.343..390F}. PAHs are expected to be photodissociated by multi-photon events in the inner disk atmospheres \citep[$<0.5$ au,][]{2007A&A...466..229V}. PAHs are consequently not included in the models and all carbon locked in PAHs is assumed to be released in gas phase. The resulting fractional elemental abundances are given in Table \ref{table:elementalabound} for dust-free models and for dusty models with $Q = 6\times 10^{-3}$. This value, which corresponds to a fractional abundance of grain of $6.9 \times 10^{-11}$ (no sublimation of the grains), is taken as the reference for dusty models and we define $Q_{\text{ref}} \equiv 6\times 10^{-3}$.

\subsection{Chemical network}

\begin{center}
\begin{figure}
\centering
\includegraphics[width=.34\textwidth]{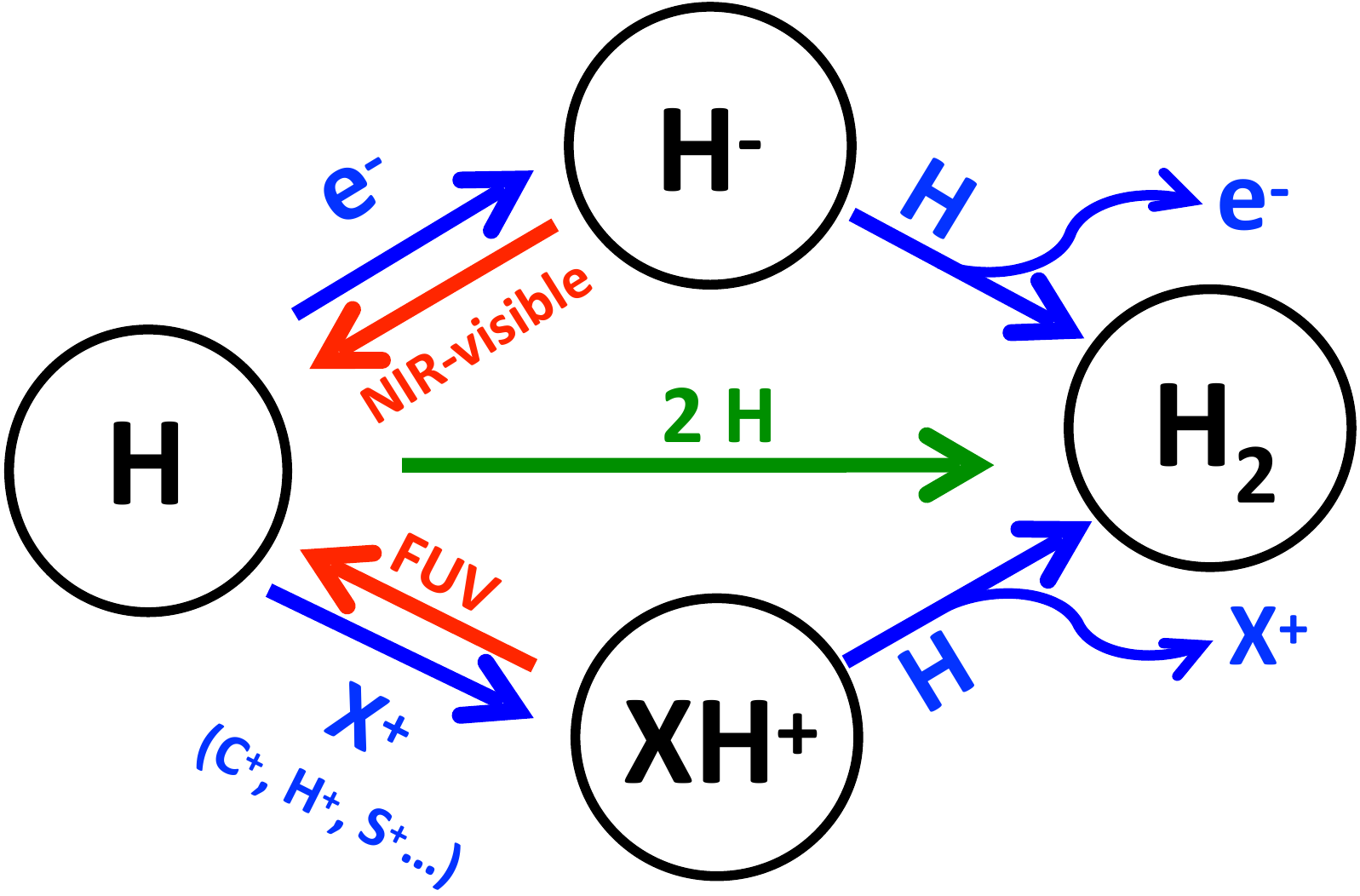} 
\caption{Dust-free formation routes for H$_2$ where X stands for H, C, S, and Si. Blue arrows represent two-body reactions, \REV{the green arrow} the three-body reaction and red arrows photodestruction of key intermediates.}
\label{fig:H2routes}
\end{figure}
\end{center}

The chemical network is constructed from \citet{2015A&A...578A..63F} and \citet{2019A&A...622A.100G}. It includes 140 species and about a thousand gas-phase reactions. Details on the chemical reactions added to the former network, together with adopted rate coefficients, are given in Appendix \ref{app:chemical-network}.

Regarding the formation of H$_2$, three gas-phase formation routes are included (see Fig.\ref{fig:H2routes}).

\begin{enumerate}
\item Electron catalysis through the intermediate anion H$^-$, via radiative attachment followed by fast associative detachment:

\begin{align}
&\Hy  + \text{e}^-  \rightarrow \Hy^- + h \nu,  \label{eq:rad-el-asso} \\
&\Hy^- + \Hy  \rightarrow \Hy_2 + \text{e}^-. \label{eq:asso-det}
\end{align}
This route has been shown to be quenched due to the photodetachment of the fragile H$^-$ by visible and NIR fields in T Tauri stellar winds \citep{1988MNRAS.230..695R, 1989ApJ...336L..29G} but efficient in the early universe at z$<$100 \citep{2013ARA&A..51..163G}. We reconsider the role of this route in Section \ref{sec:singlepoint}.\\

\item Ionic catalysis by any ion noted here X$^+$ (with X~=~H, C, S, and Si), through the intermediate ion XH$^+$ via radiative association followed by an ion-neutral reaction:
\begin{align}
&\text{X}^+ + \Hy \rightarrow \text{X}\Hy^+ + h \nu,  \label{eq:rad-asso-XH+}  \\
&\rm{X}\Hy^+ + \Hy \rightarrow \Hy_2 + \rm{X}^+ . \label{eq:rec-XH+} 
\end{align}
This route is very similar to the former though built from X$^+$ instead of e$^-$. In contrast with H$^-$, XH$^+$ ions are photodestroyed by UV photons and can thus survive the strong visible radiation fields. Surprisingly, previous models of dust-free stellar winds have never discussed formation routes via CH$^+$, SiH$^+$ or SH$^+$, focusing only on the formation route via H$_2^+$. We show below and in Appendix \ref{app:anal-H2} that the latter is inefficient compared to the formation by CH$^+$ or SiH$^+$.\\

\item Three-body reaction:
\begin{align}
\Hy + \Hy + \Hy  \rightarrow \Hy_2 + \Hy,
\end{align} 
that is relevant at high density.
\end{enumerate}
Complementary reactions involved in the chemistry of the intermediates H$_2^+$ and H$^-$ have also been included (see Appendix \ref{app:chemical-network}). 

Regarding H$_2$ formation on dust, we adopt the formation rate of \cite{1979ApJS...41..555H} assuming a single grain size distribution of radius $a_g$. An important caveat is the gas temperature dependence of the probability $S(T_{\rm{K}})$ that a H atom sticks when it collides with a grain. In the high temperature regime relevant for jets, large discrepancies exist in the literature regarding sticking probabilities, especially when including chemisorption or specific substrates \citep[see][Appendix A]{2013MNRAS.436.2143F}. Our adopted expression for $S(T_{\rm{K}})$ (see eq. (\ref{eq:stickingproba})) gives a lower limit on the formation rate of H$_2$ at high temperature.

In dust-poor gas, atomic species can dominate the opacity of the gas in the UV. When ionized, they can also contribute to the synthesis of the key anion H$^-$ by increasing the electron fraction.  
Thus, a reduced chemical network for Mg/Mg$^+$, incorporating photoionization and charge exchange reactions, has been added. Other rarer elements such as Li, Al or Na are not expected to contribute significantly in either the attenuation of the radiation field or the ionization balance of the gas and are consequently not included in the model.

\section{Chemistry}
\label{sec:singlepoint}

To examine the formation and destruction routes of the main molecules observed in protostellar jets, single spatial point models have been computed over a range of physical conditions representative of protostellar jets. The evolution of the gas is assumed to be isothermal and isochoric. Parameters of the models presented in this section are proton density $n_{\text{H}}$, kinetic temperature $T_{\rm{K}}$, unshielded FUV radiation field $G_0$ and dilution factor of the (visible) black-body radiation field at $4000$~K noted by $W$. The explored parameter space is summarized in Table \ref{table:parameter-space-single-point}. In this section, we discuss steady-state chemical abundances. This approach, though simple, allows us to identify relevant chemical reactions and capture the essential features of the chemistry operating in jets. The results of this section are summarized in Sec. \ref{subsection:summary-sec-single-point}.

\begin{table}
\caption{Physical parameters explored in single point models and their fiducial values.} 
\label{table:parameter-space-single-point} 
\begin{tabular}{l c c c}    
\hline        
Parameter & Symbol & Range or value & Fiducial \\    
\hline                                   
Density & $\nH$ & $10^{5}$ - $3 \times 10^{12}$\dens~ & $10^9$\dens \\
Temperature & $T_K$ & 200 - 5000~K& 1000K \\
Dust fraction & $Q/Q_{\text{ref}}$ & 0 - 1 & 0 \\
FUV field\tablefootmark{a} & $G_0$ & $10^4$ & $10^4$\\
Visible field\tablefootmark{a} & $T_{\rm vis}$ & 4000 K & 4000 K \\
Visible dilution\tablefootmark{a} & $W$ & $5 \times 10^{-7}$ & $5 \times 10^{-7}$ \\
\hline 
\end{tabular}
\tablefoottext{a}{The mean intensity of the radiation field is given by $J_{\nu} = W B_{\nu}(T_{\rm vis})+ J_{\nu}^{FUV}$, where $B_{\nu}(T_{\rm vis})$ is the intensity of a black-body radiation field at a temperature $T_{\rm vis}$ and $J_{\nu}^{FUV}$ the FUV part of the ISRF rescaled by $G_0$ (see Appendix \ref{app:RF})}.
\end{table}

\subsection{Formation of H$_2$}
\label{subsection:H2-routes}

\begin{center}
\begin{figure}
\centering
\includegraphics[width=.5\textwidth]{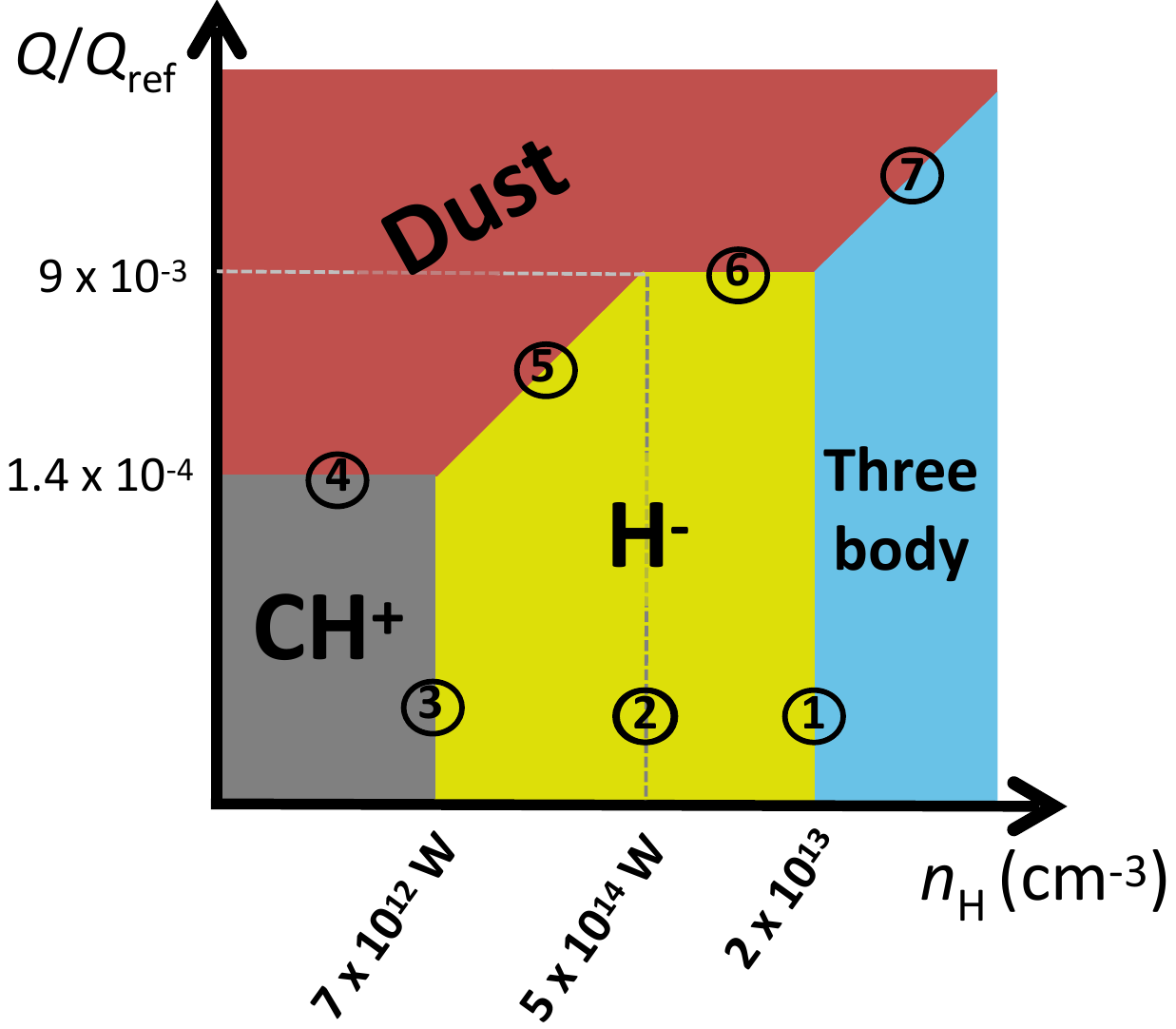}
\caption{Schematic view of the dominant H$_2$ formation routes depending on the density, the dust fraction and visible radiation field $W$ summarizing our results presented in Section \ref{subsection:H2-routes} and in Appendix \ref{app:anal-H2}. The location of the boundaries are given for a temperature of $T_{\rm{K}} = 1000$~K, $x_{\rm{e}} = 4.8~10^{-4}$, and $x(\rm{C}^+) = 3.6~10^{-4}$. The schematic view remains valid from $\simeq 100~$K up to $\simeq 5000$~K. Dependency on $T_{\rm{K}}$, $x_{\rm{e}}$, $x(\rm{C}^+)$ are given in Appendix \ref{app:anal-H2}. \REV{Note that some limits depend on the visible flux W and others do not.} Depending on the visible flux, boundaries \ding{172}, \ding{173}, and \ding{174} can merge.}
\label{fig:H2-summary}
\end{figure}
\end{center}

Formation of H$_2$ constitutes the first step of molecular synthesis. The dominant gas-phase formation route of H$_2$ and its efficiency depends mainly on the ability of H$^-$ to survive to photodetachment (see Fig. \ref{fig:H2routes}). 
The influence of dust on the chemistry depends then critically on the efficiency of  gas-phase routes. Our results are summarized in Fig. \ref{fig:H2-summary} for specific values of $T_{\rm{K}} = 1000$~K, $x_{\rm{e}} = 4.8 \times 10^{-4}$, and $x(\rm{C}^+) = 3.6 \times 10^{-4}$. We first study H$_2$ formation in the absence of dust (bottom part of the Fig. \ref{fig:H2-summary} with boundaries \ding{172}, \ding{173}, and \ding{174}) and then the influence of a non-vanishing dust fraction (bulk of the diagram with boundaries \ding{175}, \ding{176},\ding{177}, and \ding{178}). To generalize our results obtained here for a single set of values of $T_{\rm{K}}$, $G_0$ and $W$, we also propose in Appendix \ref{app:anal-H2} a numerically validated analytical approach that provides expressions for each boundary, and the associated formation rates of H$_2$ for any physical condition.

\begin{center}
\begin{figure}
\centering
\includegraphics[width=.47\textwidth]{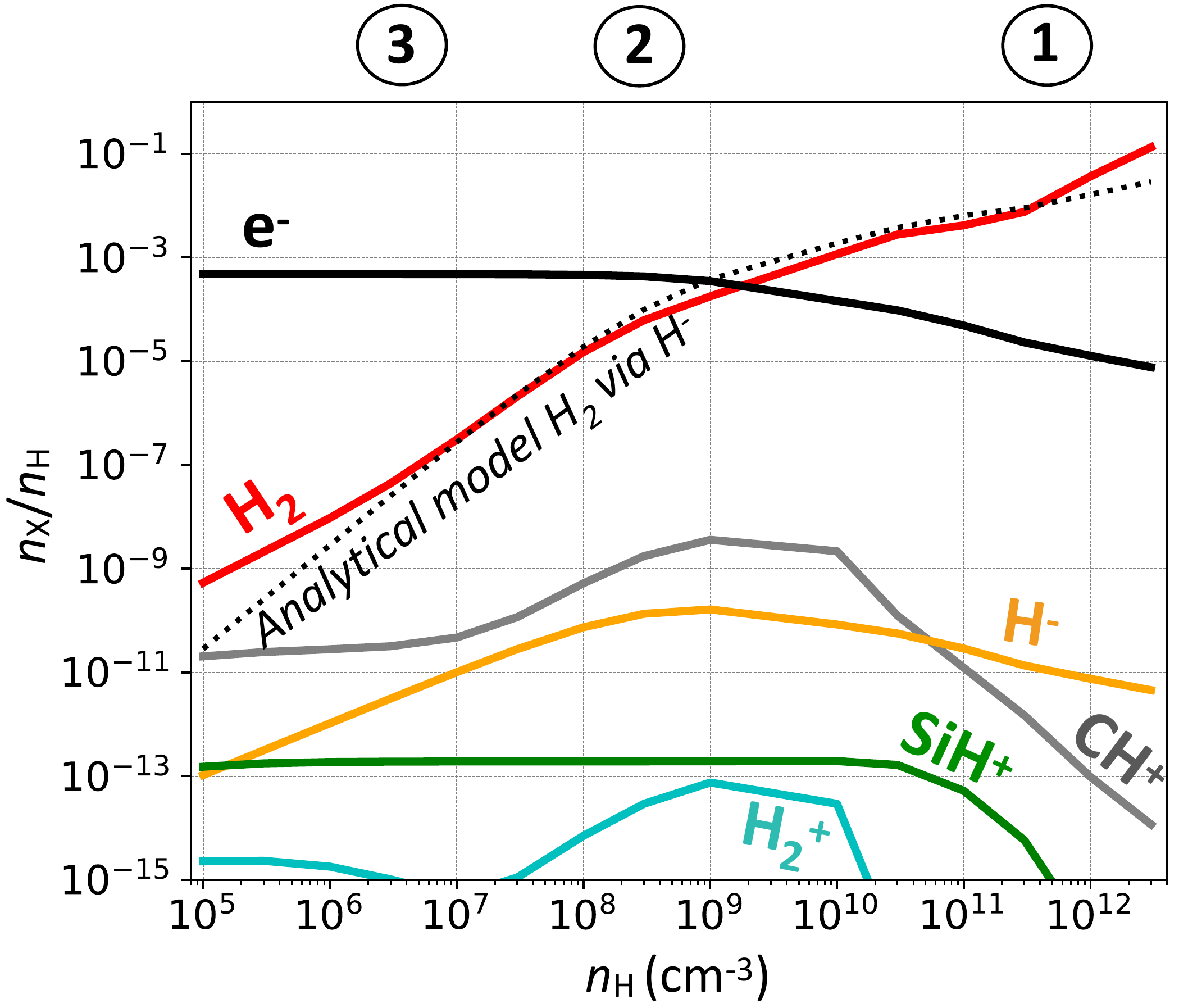} 
\caption{Steady-state abundances relative to total H nuclei for H$_2$ and chemical species involved in its formation for dust-free single point models with $G_0=10^4$, $W=5 \times 10^{-7}$, $T_{\rm{K}}=1000$ K, and $\nH$ ranging from $10^{5}$ to $2 \times 10^{12}$~cm$^{-3}$. An analytical expression of the steady state abundance of H$_2$ assuming destruction by photodissociation and formation by H$^-$ only is also ploted in black dotted line (see Appendix \ref{app:anal-H2}). Boundaries defined in Fig. \ref{fig:H2-summary} are also indicated on the upper axis. Note that because of the decrease of the electron fraction following recombination at high density, three-body reaction takes over from the formation via H$^-$ (boundary \ding{172}) at a lower density than indicated in Fig. \ref{fig:H2-summary}.}
\label{fig:H2-chemistry}
\end{figure}
\end{center}

\subsubsection{Dust-free}

Figure \ref{fig:H2-chemistry} shows that in the absence of dust, and for a radiation field characterized by $G_0=10^4$ and $W=5 \times 10^{-7}$, the gas remains atomic up to $\nH = 3 \times 10^{12}$\dens. The abundance of H$_2$ increases from $\simeq 10^{-9}$ up to $\simeq 0.2$. Over the explored range of density, H$_2$ is preferentially photodissociated by the unshielded radiation field. Since the photodissociation rate does not depend on the density, the global trend seen in H$_2$ is mostly due to the formation routes. 

We also plot an analytical model of the abundance of H$_2$ assuming formation by H$^-$ only and destruction by photodissociation (see Appendix \ref{app:anal-H2}, eq. (\ref{eq:analytic-H-})). The analytic expression reproduces very well the global increase of H$_2$ from $\nH=3 \times 10^6$ to $3 \times 10^{11}$\dens. In this regime, formation by H$^-$ is the dominant formation route of H$_2$ (boundaries \ding{174} to \ding{172}, Fig. \ref{fig:H2-summary}). The route though H$^-$ being a catalytic process by electrons, its efficiency depends linearly on the electron fraction. The recombination of ions at high density reduces the electron fraction and thus, the efficiency of this route.
Below $n_{\rm{H}} \simeq 2 \times 10^8$\dens~(boundary \ding{173}), photodetachment of H$^-$ takes over from the reaction H$^-$ + H $\rightarrow$ H$_2$ + e$^-$ leading to a decrease of H$^-$ and limiting the formation rate of H$_2$. Our analytical approach generalizes this result to any density and radiation field and shows that for a diluted black-body at 4000~K, this transition appears for (boundary \ding{173} and Appendix \ref{app:anal-H2}, eq. (\ref{eq:boundary-2}))
\begin{equation}
\frac{n_H}{W} = 4.6 \times 10^{14}~ \rm{cm}^{-3}.
\label{eq:def-strong-visible}
\end{equation}
Despite the photodestruction of H$^-$, formation by H$^-$ remains the dominant formation route of H$_2$ down to $\nH = 3 \times 10^6$\dens.

Below this value (boundary \ding{174}), the analytical model underestimates the H$_2$ abundance. In this regime, formation via CH$^+$ takes over from formation by H$^-$. This is due to a quenching of H$^-$ route caused by a rapid photodetachment of H$^-$. For example, at the boundary \ding{174}, only $\sim 1\%$ of H$^-$ formed by radiative attachment is actually converted in H$_2$. More generally, this transition appears for (boundary \ding{174}, Appendix \ref{app:anal-H2}, eq. (\ref{eq:boundary-3}))
\begin{equation}
n_{\text{H}}/W = 6.7 \times 10^{12} \text{cm}^{-3} \left(\frac{x(\text{C}^+)}{3.6~10^{-4}}\right) \left(\frac{x_e}{4.8~10^{-4}}\right)^{-1} \left( \frac{T_{\text{K}}}{1000~\text{K}} \right)^{-1.32},
\end{equation}
where $x(\rm{C}^+)$ is the abundance of C$^+$. Interestingly, the very low radiative association rate of S$^+$ with H prevents this already rare element to participate significantly to the formation of H$_2$ via SH$^+$. Formation via H$_2^+$ is found to be negligible over the explored parameter range. As seen in Fig. \ref{fig:H2-chemistry}, H$_2^+$ is always at least two orders of magnitude less abundant than H$^-$, CH$^+$ or SiH$^+$ and does not form H$_2$ at similar levels.

Above $\nH = 3 \times 10^{11}$\dens, the analytic model also underestimates the abundance of H$_2$. In this regime, the three-body reaction route takes over from the formation by H$^-$. When H$^-$ is not photodetached, this transition appears for (boundary \ding{172} and Appendix \ref{app:anal-H2}, eq. (\ref{eq:boundary-1}))
\begin{equation}
n_{\text{H}} = 1.9 \times 10^{13} ~\text{cm}^{-3}~ \left( \frac{x_e}{4.8~10^{-4}}  \right) \left( \frac{T_{\text{K}}}{1000~\text{K}} \right)^{1.24}.
\label{eq:ncritH3body}
\end{equation}

\subsubsection{Dust-poor}
\label{subsec:chem-dust-H2}

\label{subsection:H2-dust}

\begin{center}
\begin{figure}
\centering
\includegraphics[width=.48\textwidth]{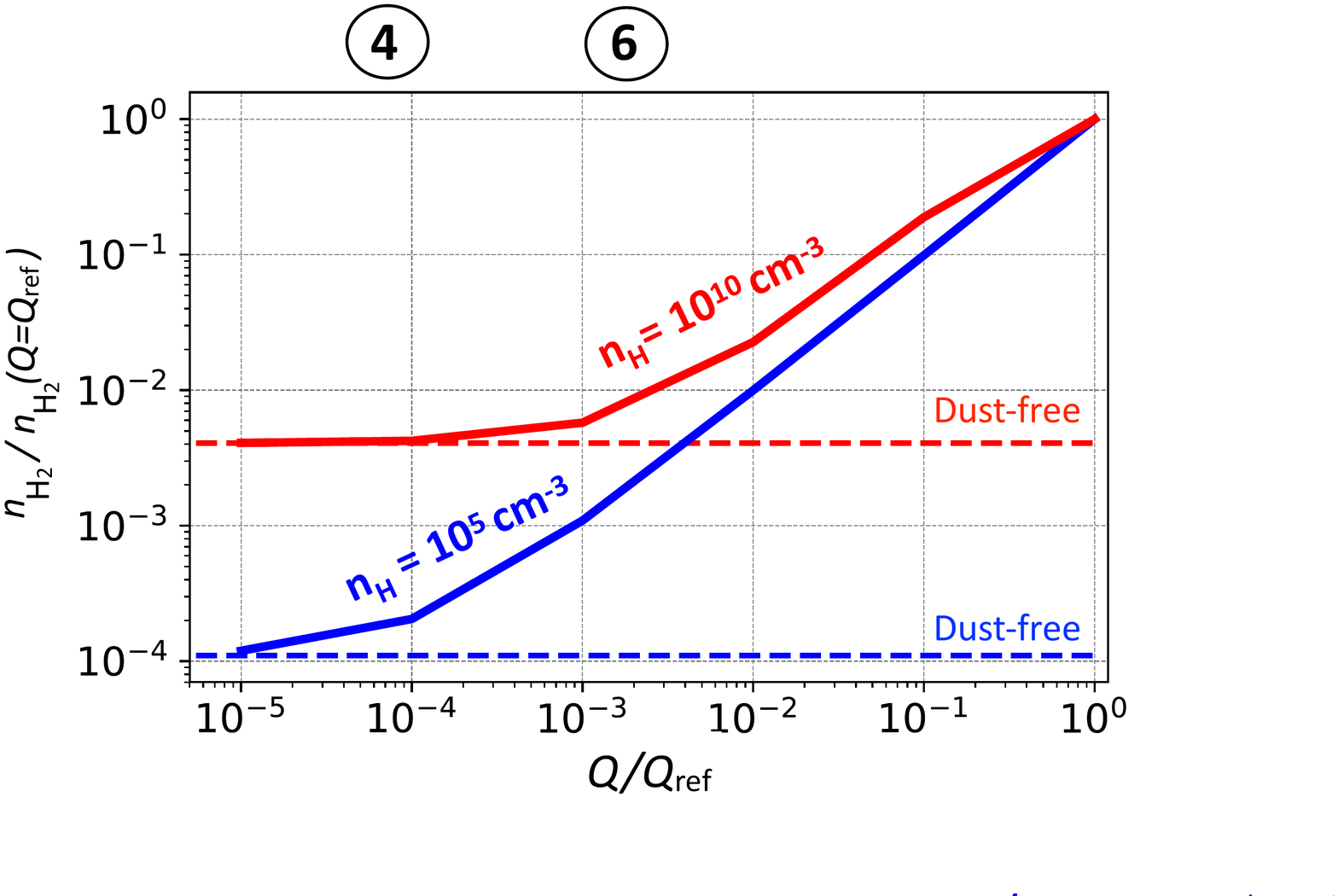}
\caption{Steady-state abundances of H$_2$ normalized to its value at $Q/Q_{\text{ref}} = 1$ as a function of $Q/Q_{\text{ref}}$ for two densities : $\nH=10^{10}$\dens~(solid red line) and $\nH=10^{5}$\dens~(solid blue line). Other parameters are constant and equal to their fiducial values (see Table \ref{table:parameter-space-single-point}). Gas-phase formation of H$_2$ is dominated by H$^-$ for $\nH=10^{10}$\dens~($\nH/W=2\times 10^{16}$\dens) and by CH$^+$ for $\nH=10^{5}$\dens~($\nH/W=2\times 10^{11}$\dens). Dashed lines indicate H$_2$ abundance in the absence of dust. Abundances for each set of model are normalized to their value at $Q/Q_{\text{ref}}=1$. Note that because of the low electron fraction at $\nH=10^{10}$~cm$^{-3}$ ($x_e \simeq 10^{-4}$) formation on grains takes over from the formation via H$^-$ (boundary \ding{177}) at a lower dust fraction than indicated in Fig. \ref{fig:H2-summary}.}
\label{fig:H2-dust}
\end{figure}
\end{center}

Since these local models assume no attenuating material to the source, the inclusion of dust does not affect significantly the efficiency of gas-phase formation routes of H$_2$. Consequently, our previous results on dust-free chemistry still hold. The effect of increasing dust fraction $Q/Q_{\text{ref}}$ is to add a new formation route that can compete with gas-phase formation routes. The critical amount of dust above which formation on dust grain takes over depends on the efficiency of the dust-free formation route.    
Figure \ref{fig:H2-dust} shows the variation of the abundance of H$_2$ as function of the dust fraction $Q/Q_{\text{ref}}$ for the fiducial radiation field, and for $\nH= 10^{10}$\dens~and $\nH= 10^{5}$\dens. As shown above, gas-phase formation routes are dominated by H$^-$ in the first case and by CH$^+$ in the latter.

Below $Q/Q_{\text{ref}} \simeq 10^{-3}$ and for $\nH = 10^{10}$\dens, the H$_2$ abundance is independent of the gas-to-dust ratio $Q$ and equal to its dust-free value. In this regime, H$_2$ is formed through H$^-$ with a maximal efficiency (H$^-$ is not photodetached) and formation on dust is negligible. At about $Q/Q_{\text{ref}} \simeq 2 \times 10^{-3}$ formation on dust takes over from gas-phase formation, driving up the H$_2$ abundance.
A more detailed analysis (see Appendix \ref{app:anal-H2}) shows that when H$^-$ is the main formation route and not photodetached, the critical dust fraction below which H$_2$ gas-phase formation takes over from formation on grains is (Fig. \ref{fig:H2-summary}, boundary \ding{177})
\begin{equation}
Q/Q_{\text{ref}} = 9.3 \times 10^{-3} \left( \frac{T_{\rm{K}}}{1000~\text{K}} \right)^{0.4} \left( \frac{S(T_{\rm{K}})}{S(1000~\text{K})} \right)^{-1} \left( \frac{x_e}{4.8~10^{-4}} \right),
\label{eq:ncritH3body}
\end{equation}
where $S(T_{\rm{K}})$ is the sticking coefficient of H on grains adopted from \citet{1979ApJS...41..555H}. When the formation through H$^-$ is reduced by photodetachment, this critical dust fraction is proportional to $n_{\rm{H}}/W$ (Fig. \ref{fig:H2-summary}, boundary \ding{176} and Appendix \ref{app:anal-H2} eq. (\ref{eq:boundary-56})).

Figure \ref{fig:H2-dust} shows that for a lower density-to-visible field ratio ($\nH/W=2 \times 10^{11}$\dens) dust has a significant impact at a much smaller dust fraction. As seen in the previous section, the gas-phase formation of H$_2$ is then dominated by CH$^+$. Formation by CH$^+$ being about two orders of magnitude less efficient than electron catalysis, the critical dust fraction above which formation on grains takes over from gas-phase formation is accordingly lowered by a similar factor (Fig. \ref{fig:H2-summary}, boundary \ding{175} and Appendix \ref{app:anal-H2} eq. (\ref{eq:boundary-4})).

\subsection{Other molecules}
\label{subsec:single-point-molecule}

H$_2$, even when scarce, constitutes the precursor of other molecules such as CO, SiO, OH, and H$_2$O that are observed in protostellar jets. The molecular richness depends then on the abundance of H$_2$. As such, the inclusion of dust increases molecular abundances only by increasing the fraction of H$_2$. However, other physical variables can regulate molecular abundances such as the temperature, the FUV radiation field and the density. Since molecules are essentially formed by two-body reactions and destroyed by photodissociation in the UV domain, molecular abundances depends mostly on the ratio $\nH/G_0$. In this section, results on molecular abundances obtained by varying $\nH$ for fixed $G_0 = 10^4$ can thus be generalized for any $G_0$ by translating to the ratio $\nH/G_0$.

\subsubsection{Dust-free}

\begin{center}
\begin{figure}
\centering
\includegraphics[width=.5\textwidth]{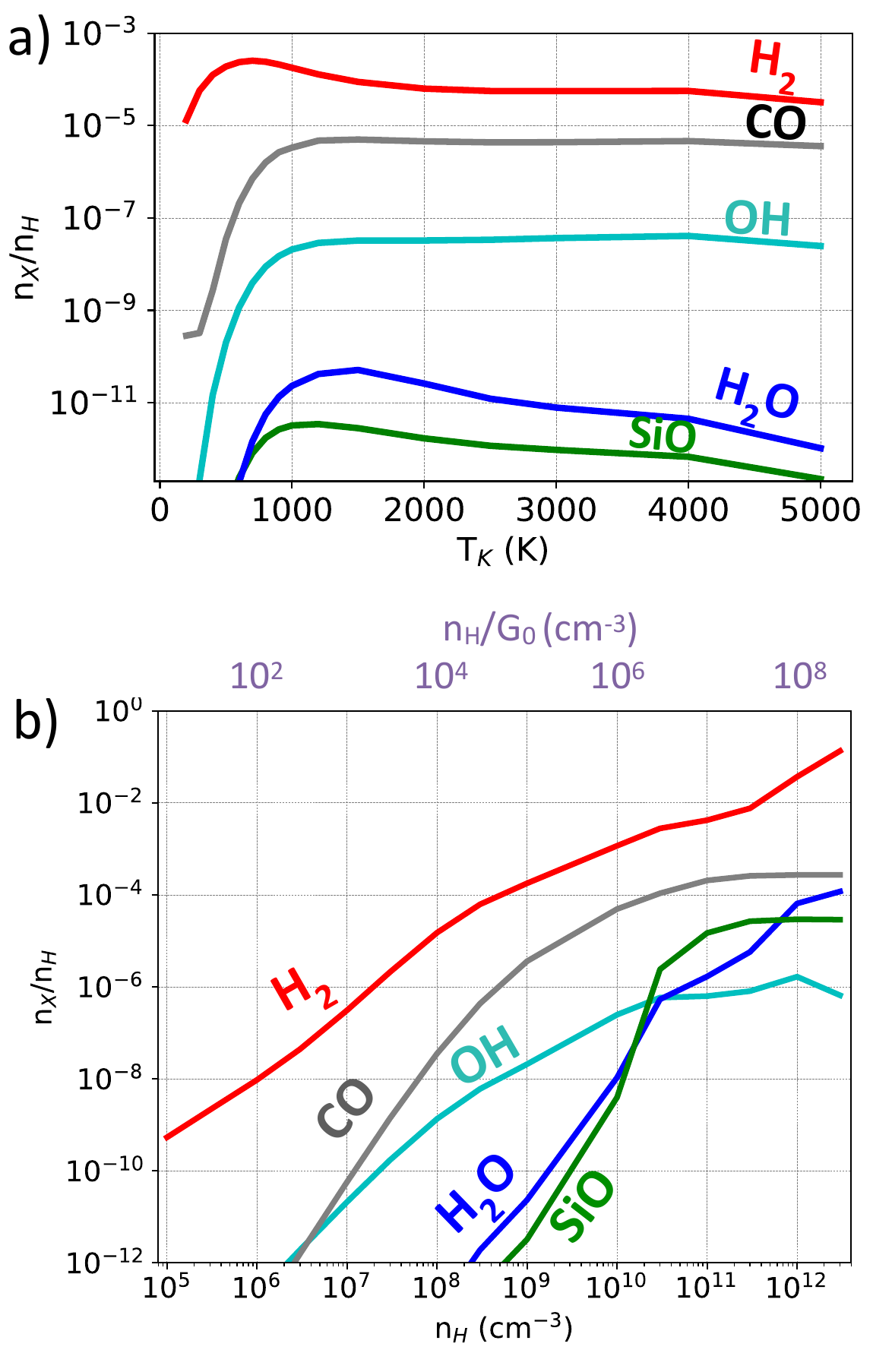}
\caption{Steady-state abundances relative to total H nuclei for relevant molecular species from single point models in the absence of dust. a) Abundances as function of temperature for $G_0=10^4$, $W=5 \times 10^{-7}$, and $\nH = 10^{9}$\dens, b) abundances as function of $\nH$ (lower axis) and $\nH/G_0$ (upper axis) for $T_{\rm{K}} = 1000$~K, $G_0=10^4$, and $W=5 \times 10^{-7}$.}
\label{fig:molec-chemistry}
\end{figure}
\end{center}

Figure \ref{fig:molec-chemistry}-a shows that steady-state abundances of OH, CO, H$_2$O and SiO increase by several orders of magnitude with temperature and reach maximum abundances above $\simeq 1000$~K.
This trend is due to the activation of endothermic gas-phase formation routes at high temperature. Formation of CO, SiO and H$_2$O is indeed initiated by the formation of OH through the neutral-neutral reaction
\begin{equation}
\text{O} + \text{H}_2 \rightarrow \text{OH} + \text{H} ~~~~~~~~~~ \Delta E = +2980~\text{K}.
\label{eq:reaction-OH}
\end{equation}
This warm route involving H$_2$ is fundamental for the formation of all the considered molecules, even if the H$_2$ fraction is low.

\begin{center}
\begin{figure}
\centering
\includegraphics[width=.5\textwidth]{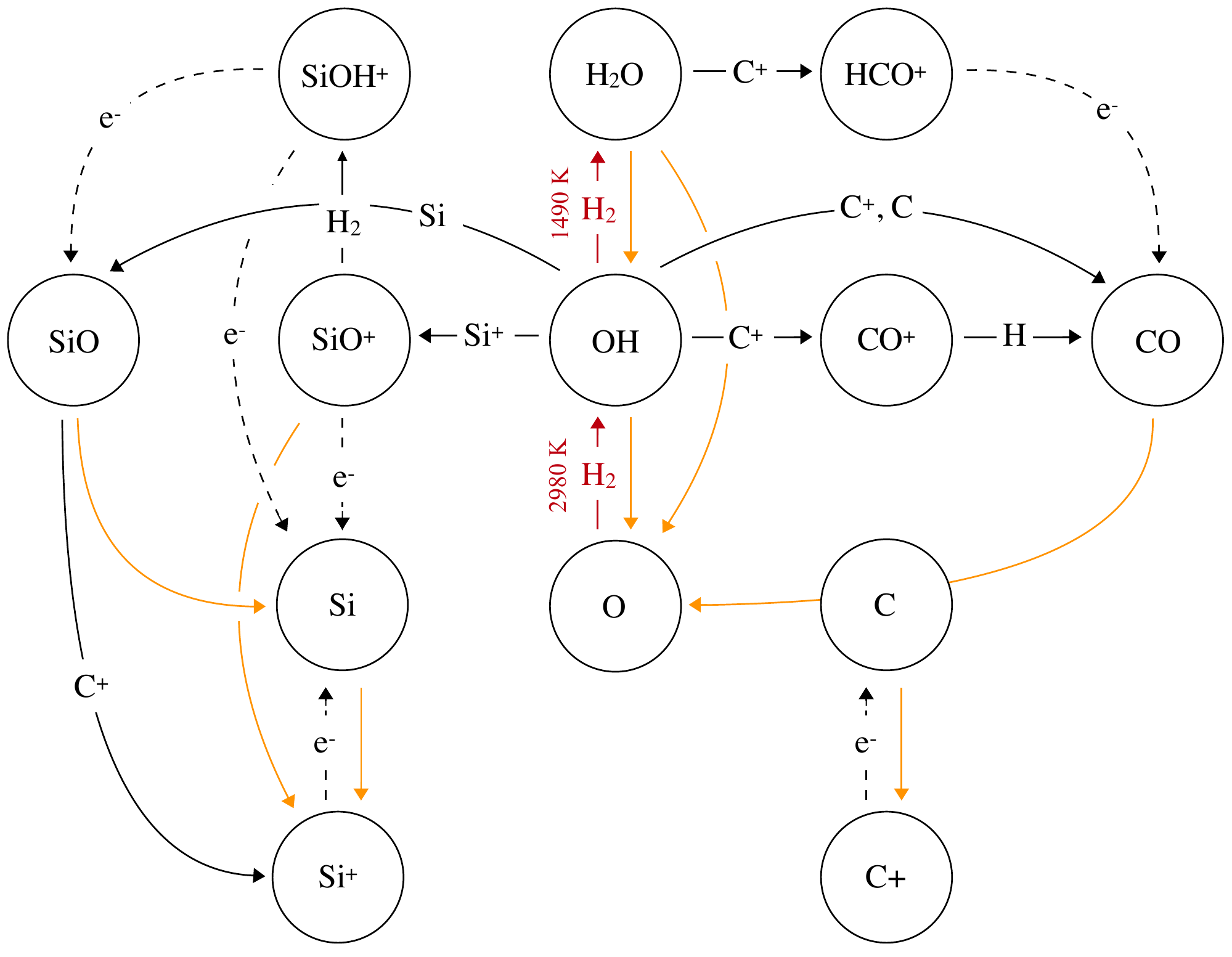}
\caption{Dominant reactions controlling the abundance of CO, H$_2$O, and SiO under warm ($ T_{\rm{K}} \ge 800$~K) and irradiated conditions. OH appears to be a key intermediate for the three species. The ionization state of carbon controls the destruction of SiO  and H$_2$O, and the formation of CO. The ionization state of silicon controls the formation of SiO. For unshielded ISRF FUV radiation field, carbon and silicon are ionized for $n_{\rm{H}}/G_0 \le 10^5$\dens~and  $n_{\rm{H}}/G_0 \le 3 \times 10^{6}$\dens, respectively.}
\label{fig:molecule-formation-scheme}
\end{figure}
\end{center}

Figure \ref{fig:molec-chemistry}-b shows steady-state abundances as function of density for a temperature of $T_{\rm{K}} = 1000$~K, sufficient for efficient molecule formation. As H$_2$ and OH increase with $\nH$, the abundances of the other species increase as well. 
For $\nH \ge 3 \times 10^{10}$\dens, the gas is essentially atomic but rich in CO, SiO, H$_2$O. This is one of the most fundamental and unique characteristics of dust-free chemistry. Still, due to complex formation and destruction pathways, each molecular species behaves differently as a function of $\nH$, revealing their formation and destruction routes in H$_2$ poor gas. Figure \ref{fig:molecule-formation-scheme} summarizes the dominant reactions contributing to the formation and destruction of CO, H$_2$O, SiO depending on the ionization state of C and Si. 

The CO formation pathway is essentially regulated by the ionization state of the carbon. Below $\nH = 2 \times 10^9$\dens, carbon is ionized and CO is produced via the ion-neutral reaction C$^+$ + OH producing either directly CO or CO$^+$. In the latter, CO$^+$ is neutralised by a fast charge exchange with H. Above $\nH = 2 \times 10^9$\dens, CO is produced directly by the neutral-neutral reaction C+OH $\rightarrow$ CO. Destruction is mostly via photodissociation.

The H$_2$O abundance exhibits a stiff increase with $\nH$. Over the explored parameter range, H$_2$O is produced through the neutral-neutral reaction OH + H$_2$. Below $\nH \simeq 10^{10}$\dens, destruction is via photodissociation, and by C$^+$. 
At higher density, the main destruction route is via the reverse reaction H$_2$O + H $\rightarrow$ OH + H$_2$.

The SiO abundance exhibits a stiff increase around $\nH\simeq 3 \times  10^{10}$\dens, rising by more than four orders of magnitudes in one decade of $\nH$. This feature is due to a change in both destruction and formation routes. Below $\nH = 3 \times 10^{10}$\dens, Si$^+$ is the main silicon carrier and SiO synthesis is initiated by the ion-neutral reaction Si$^+$ + OH $\rightarrow$ SiO$^+$. However, in contrast to the analogous reaction with CO$^+$, SiO$^+$ cannot be neutralized through a charge exchange with the main collider, namely H. Alternatively, SiO$^+$ decays toward SiO though SiO$^+$ + H$_2 \rightarrow$ SiOH$^+$, eventually leading to SiO. H$_2$ being rare in the absence of dust, this formation route is much less efficient than the analogous reaction that forms CO. In addition, at low $\nH/G_0$, the abundant C$^+$ destroys efficiently SiO to form CO. Consequently when carbon and silicon are ionized, the medium is hostile to the formation and the survival of SiO. On the contrary, above $\nH \simeq  3 \times  10^{10}$\dens, SiO is formed directly through the neutral-neutral reaction Si + OH and destruction by C$^+$ is quenched by the recombination of carbon. In contrast with the analogous reaction with H$_2$O, the reverse reaction SiO + H has a very high endothermicity ($3.84$~eV) that prevents any destruction by H. Given these favorable factors, SiO becomes the main silicon reservoir above $\nH \simeq 3 \times 10^{10}$\dens.

\subsubsection{Dust-poor}
The inclusion of dust increases the molecular abundances by increasing the H$_2$ abundance. Thus, the minimal amount of dust required to affect the chemistry of CO, OH, SiO, H$_2$O is similar to that determined in the Sec. \ref{subsec:chem-dust-H2}.  Molecular abundances are then increased accordingly but specific formation and destruction routes remain the same. 
 
\subsection{Summary}
\label{subsection:summary-sec-single-point}

In this section, the chemistry of dust-free and dust-poor gas has been studied. Regarding H$_2$, we show that the dominant gas-phase formation route and its efficiency depends
critically on the ratio between the density and the visible radiation field, quantified here by $\nH/W$. Above $\nH/W  \simeq 5 \times 10^{14}$\dens, H$_2$ is efficiently formed via H$^-$ whereas below this value, photodetachment reduces its efficiency. 
When optimally formed via H$^-$, a minimum fraction of dust of about $Q/Q_{\text{ref}} \simeq 2\times 10^{-3}$ is required to have a significant impact on the chemistry.

Regarding CO, OH, SiO and H$_2$O, we find that high abundances are reached for $n_{\rm{H}}/G_0 \ge 10^6$\dens, despite low abundances of H$_2$. Efficient formation routes are initiated by OH and require a warm environment ($T_{\rm{K}} \ge 800$~K). The inclusion of dust increases molecular abundances by increasing the H$_2$ abundance accordingly. We also find that the abundance of SiO is very sensitive to the ionization state of carbon and silicon. When both species are singly ionized, the SiO abundance is very low due to both destruction by C$^+$ and a very low efficiency of the formation by Si$^+$ in an H$_2$-poor environment.

\section{1D wind models}
\label{sec:wind}


In this section, the chemistry studied in the previous section is incorporated in a more 
realistic model of 1D wind \SC{streamline}. \SC{In addition to the quantities found to 
control the chemistry in unattenuated static environments (namely $\nH/G_0$, $\nH/W$, $T_{\rm K}$, 
$Q/Q_{\text{ref}}$, and $T_{\text{vis}}$),} we include three ingredients: the attenuation 
of the radiation field with the distance from the source, the time-dependent chemistry, 
and the differential geometrical dilutions of the density and the radiation field. \SC{We present
below our simple 1D model where these effects are implemented with a minimal number of free 
parameters, allowing to investigate a wide range of source evolutionary phases.}

\begin{figure*}
\centering
\includegraphics[width=.20\textwidth]{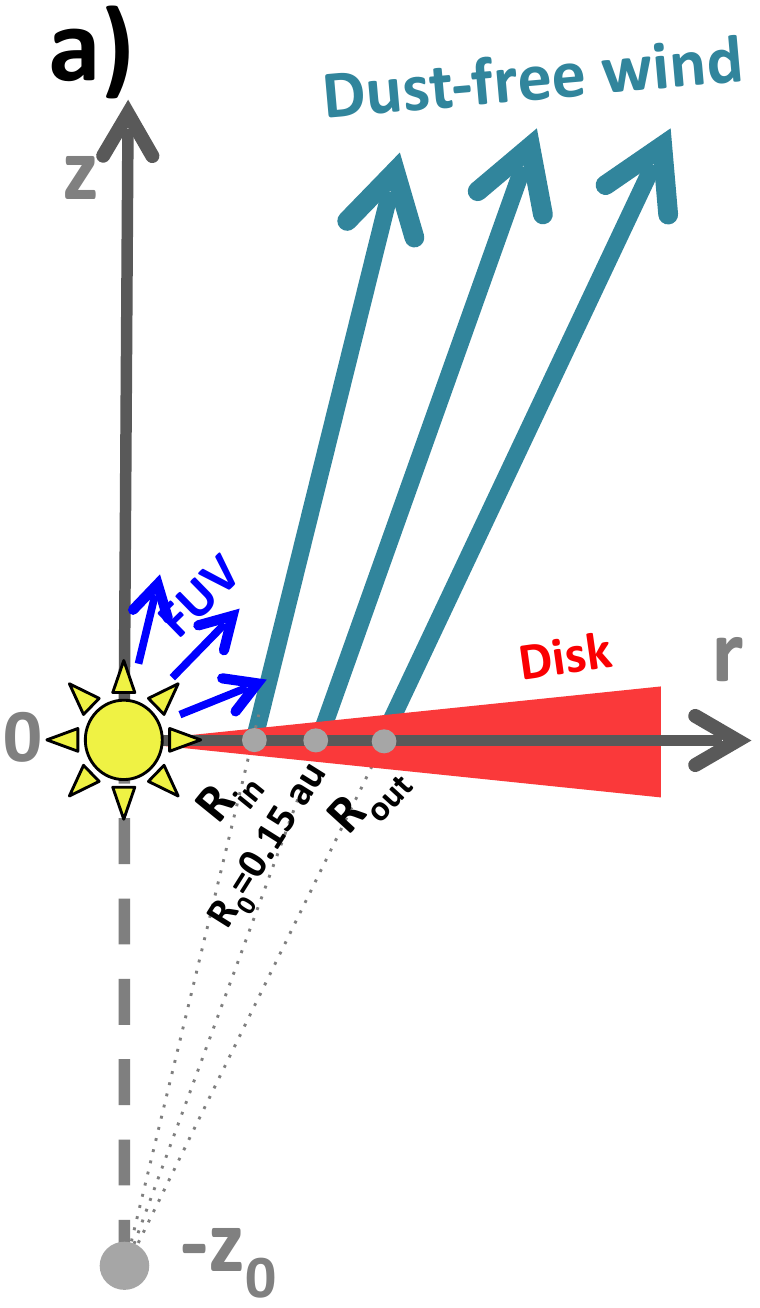}
\includegraphics[width=.79\textwidth]{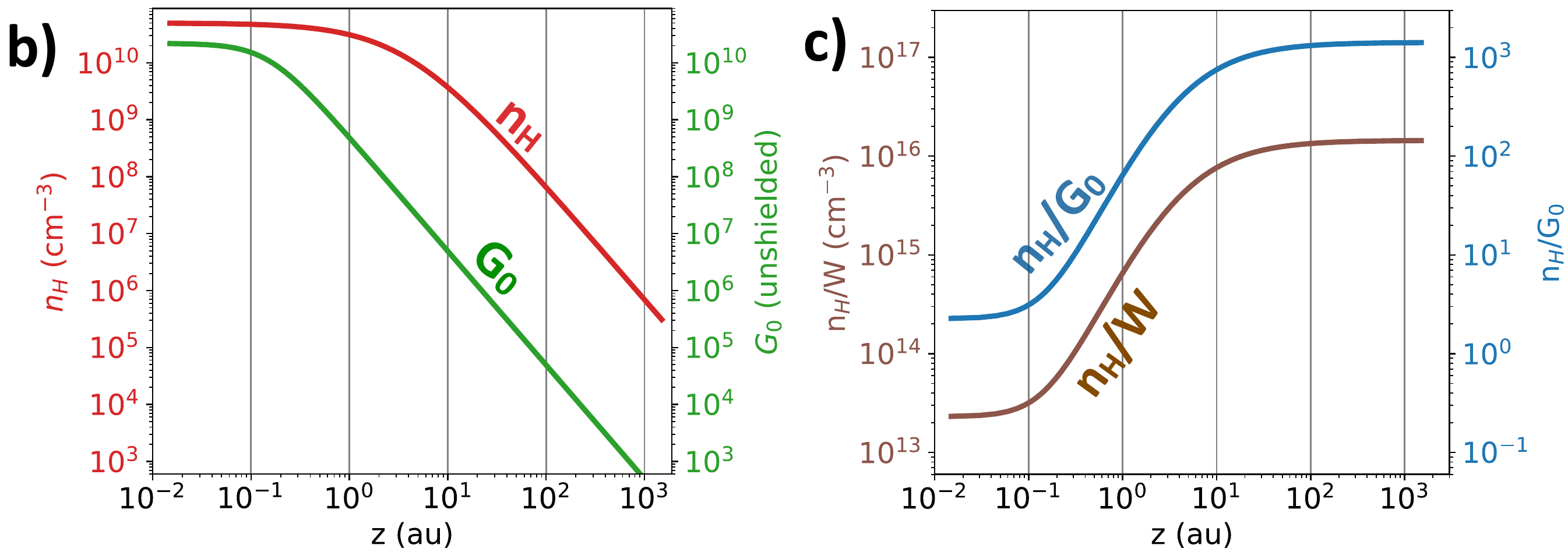}
\caption{Wind model adopted in this work. a) Schematic view of the geometry of the model. Streamlines are assumed to be straight lines launched from the disk (see Fig. \ref{fig:shematic-view}-a). The wind velocity $V_j$ is constant and equal to 100~\kms. The wind launching region extends from $R_{\text{in}} = 0.05~$au out to $R_{\text{out}} = 0.3$~au.  We focus on the chemical evolution of a representative streamline launched from 0.15~au in the disk and reduce the problem to 1D (see Sec. \ref{subsec:themodel}). b) Prescribed density $n_{\rm{H}}$ and unshielded FUV flux  $G_0$ profile along the representative streamline launched from 0.15~au for wind solution \textcircled{a} (see Table \ref{table:wind-models}). For other models,  $n_{\rm{H}}$ and $G_0$ are simply rescaled according to the eqs. (\ref{eq:density-base}) and (\ref{eq:G0-base}), respectively. c) Prescribed $n_{\rm{H}}/G_0$ and $n_{\rm{H}}/W$ ratio along the same streamline. Note that due to the collimation of the flow ($z_0 \gg R_0$), these ratios are increasing with distance by a factor $(z_0/R_0)^2 = 625$. In the absence of dust, the opacity of the gas in the visible is negligible so that $n_{\rm{H}}/W$ ratio is expected to be the true local ratio between density and visible field.}
\label{fig:wind-model-prescribed}
\end{figure*}

\subsection{The streamline model}
\label{subsec:themodel}

\begin{table*}[h!]
\centering
\caption{Models explored in this work in order of increasing source age.}
\label{table:wind-models} 
\begin{tabular}{c | c | c c c | c c c c c}
\hline 
Model    & label           & $M_*$       & $\dot{M}_{\text{acc}}$   & $\dot{M}_{\text{w}}$     & $n_{\rm{H}}^0$ & $G_0^0$        & $n_{\rm{H}}^0$/$G_0^0$ & $n_{\rm{H}}^0$/$W^0$ \\
         &                 & M$_{\odot}$ & M$_{\odot}~\rm{yr}^{-1}$ & M$_{\odot}~\rm{yr}^{-1}$ & cm$^{-3}$      &                & cm$^{-3}$              & cm$^{-3}$            \\
\hline
         &                 & 0.1         & 5(-5)                    & 5(-6)                    & 1.3(11)        & 5.5(10)        & 2.4                    & 5.9(13)              \\
Class 0  & \textcircled{a} & $\bf{0.1}$  & $\bf{2(-5)}$             & $\bf{2(-6)}$             & $\bf{5.0(10)}$ & $\bf{2.2(10)}$ & $\bf{2.3}$             & $\bf{2.3(13)}$       \\
         & \textcircled{b} & $\bf{0.1}$  & $\bf{1(-5)}$             & $\bf{1(-6)}$             & $\bf{2.5(10)}$ & $\bf{1.1(10)}$ & $\bf{2.3}$             & $\bf{1.1(13)}$       \\
         &                 & 0.1         & $5(-6)$                  & $5(-7)$                  & 1.3(10)        & 5.5(09)        & 2.4                    & 5.9(12)              \\
\hline
Class I  & \textcircled{c} & $\bf{0.5}$  & $\bf{1(-6)}$             & $\bf{1(-7)}$             & $\bf{2.5(09)}$ & $\bf{5.5(09)}$ & $\bf{0.45}$            & $\bf{1.1(12)}$       \\
\hline
Class II &                 & $0.5$       & $1(-7)$                  & $1(-8)$                  & 2.5(08)        & 5.5(08)        & 0.45                   & 1.1(11)              \\
\hline 
\end{tabular}
\tablefoot{
Except for $M_*$ and $\dot{M}_{\text{w}}$, all parameters have fixed values, namely
$R_\star = 3$~R$_{\odot}$,
$T_{\rm vis} = 4000$~K, 
$V_j = 100$~km s$^{-1}$,
$T_{\rm K} = 1000$ K,
$R_{\rm in} = 0.05$~au,
$R_{\rm out} = 0.3$~au,
$R_0 = 0.15$~au, and
$z_0/R_0=25$.
Accretion to ejection rate ratio is assumed to be 0.1 so that 
$\dot{M}_{\text{w}} = 0.1 \dot{M}_{\text{acc}}$ for all models.
Models in bold are studied in more details in section \ref{subsec:wind-standard} 
(see also Fig. \ref{fig:wind-models-abund}), and are identified by a label.
The last four columns are the density, the FUV radiation field, and the ratios
$n_{\rm{H}}^0$/$G_0^0$, and $n_{\rm{H}}^0$/$W^0$ at the base of the computed streamline. 
Numbers in parentheses are powers of 10.} 
\end{table*}

The disk wind model, \SC{illustrated in Figure \ref{fig:wind-model-prescribed}-a}, is built from a simple flow geometry 
that captures the essential properties of MHD disk wind models in a parametric approach, 
without relying on a peculiar wind solution. Following \citet{2006MNRAS.370..580K} we 
assume that the wind is launched from a region of the disk between $R_{\text{in}}$ and 
$R_{\text{out}}$, and propagates along straight streamlines diverging from a point located 
at a distance $-z_0$ below the central object (Fig. \ref{fig:wind-model-prescribed}-a). 

We follow the evolution of only one representative streamline launched from $R_0$ using 
the astrochemical model presented in Section \ref{sec:intro-model}. \SC{To reduce the 
number of free parameters}, the wind velocity, noted $V_j$, is assumed to be constant 
with distance. 
Conservation of mass then yields a geometrical dilution of density along the streamline anchored at $R_0$ of
\begin{equation}
\nH(z) =  n_{H}^0  \frac{1}{\left(1+\frac{z}{z_0} \right)^{2}},
\label{eq:nH-dilution}
\end{equation}
where we note $n_{H}^0$ the density at the base  ($z = 0$). 

\SC{The radiation field is assumed to be emitted isotropically from the star position.}
Along a given streamline, it is reduced by a geometrical dilution factor, and attenuated by gas-phase species and dust, if any. To keep the problem tractable \SC{in 1D}, the attenuation of the radiation field is assumed 
to proceed along each streamline.
\SC{The geometrically diluted, unattenuated FUV field at position $z$ is given by
\begin{equation}
G_0(z) =  G_0^0  \frac{1}{\left(1+\frac{z}{z_0}\right)^2 + \left(\frac{z}{R_0}\right)^2},
\label{eq:rad-dilution}
\end{equation}
where we note $G_0^0$ the unattenuated FUV field at the base ($z = 0$). The same geometrical 
dilution applies to the unattenuated visible radiation field.} 

\SC{Figure \ref{fig:wind-model-prescribed}-b,c show that under this simple wind geometry, the radiation field is diluted 
on a spatial scale $\simeq R_0$ while the density field is diluted on a scale $z_0$.} Due 
to the collimation of the flow ($z_0 \gg R_0$), the ratios $n_H / G_0$ and $n_H/W$ (which 
would determine the flow chemistry in the absence of attenuation and time-dependent effects) 
increase with distance \SC{up to a} factor $1+(z_0 / R_0)^2$ from their initial values at 
$z = 0$. Hence, the differential dilutions of the radiation and density fields in our model 
is regulated by the wind collimation angle $(z_0 / R_0)$.

\SC{Our 1D chemical wind-model is thus controlled in principle by 9 free parameters: the 
same six parameters as in Section 3, namely $n_H^0$, $G_0^0$, $W^0$, $T_{\rm K}$, 
$Q/Q_{\text{ref}}$, and $T_{\rm vis}$ ; and three parameters controlling the wind 
attenuation, dilution, and non-equilibrium effects: the opening angle of the streamline 
$z_0/R_0$, the typical attenuation scale $z_0$ (or alternatively the anchor radius $R_0$), 
and the velocity of the wind $V_{\text{j}}$. Since we cannot explore the full parameter 
space in the present study, we chose here to fix most of them to representative values 
observed  in molecular jets and young protostars, and focus on varying the source evolutionary 
status. Namely, we take a fixed launch radius $R_0$ = 0.15 au \citep[as estimated for the SiO jet 
in HH~212 by][]{2017A&A...607L...6T}.} The ratio $z_0/R_0$ is fixed to 25, leading to an opening 
angle of the computed streamline of $5^{\circ}$, in line with the observed universal 
collimation properties of jets across ages \citep{2007LNP...723...21C,2007A&A...468L..29C}. 
The wind velocity $V_j$ is taken equal to 100~\kms, in line with typical velocities 
measured from proper motions toward Class 0 molecular jets \citep[eg.][]{2015ApJ...805..186L}. 
\SC{The flow temperature is taken as $T_{\rm K}$ = 1000~K (see discussion in Section 
\ref{subsec:discuss-shocks}).}

The radiation field emitted by the accreting protostar is modelled as a black-body of 
photospheric origin with $T_{\rm vis}= 4000$~K, plus a FUV component coming from the 
accretion shock onto the stellar surface. At the base of the streamline, the dilution 
factor of the stellar 
black-body, defined as $W \equiv \frac{J_{\nu}}{B_{\nu}}$, is (assuming $R_0 \gg$ $R_*$)
\begin{equation}
W^0 = \frac{1}{4} \left(\frac{R_*}{R_0}\right)^{2} = 2.2~10^{-3} \left(\frac{R_0}{0.15~\text{au}}\right)^{-2} \left(\frac{R_\star}{3 R_\odot}\right)^{2}.
\label{eq:W-base}
\end{equation}
The radius of the protostar is fixed to $R_\star = 3 R_{\odot}$, with $R_{\odot}$ the solar radius. 
Regarding UV excess, FUV observations of BP Tau and TW Hya show that an ISRF provides a good proxy 
for the shape of the radiation field, though neglecting the line contribution to the FUV flux 
\citep[$>35\%$,][]{2003ApJ...591L.159B}. Here, we assume that the FUV excess follows a Mathis 
radiation field \citep[][see appendix \ref{app:RF}]{1983A&A...128..212M} and neglect the UV 
line emission. For BP Tau, \citet{2003ApJ...591L.159B} find that a FUV flux of G$_0 = 560$ at 
$100$~au is required to match the FUV continuum level. Assuming that the FUV flux scales 
with the accretion luminosity $L_{\text{acc}} = \frac{G M_* \dot{M}_{\text{acc}}}{R_*}$ 
\SC{(which is $0.24 L_{\odot}$ for} BP Tau), the scaling factor at the base of the streamline 
anchored at $R_0$ is
\begin{equation}
G_0^0 = 1.1 \times 10^{10} \left(\frac{\dot{M}_{\text{acc}}}{10^{-5}M_{\odot}~\rm{yr}^{-1}}\right)  \left(\frac{M_\star}{0.1 M_{\odot}}\right) 
 \left( \frac{R_\star}{3 R_{\odot}}  \right)^{-1}  \left( \frac{R_0}{0.15~\text{au}}  \right)^{-2} ,
\label{eq:G0-base}
\end{equation}
where $\dot{M}_{\text{acc}}$ is the accretion rate onto the protostar and $M_\star$ is the mass of the protostar. 

In order to relate the density at the launching point of a streamline $n_{H}^0$ to the mass-loss 
rate of the wind $\dot{M}_{\text{w}}$, we assume that between $R_{\text{in}}$ and $R_{\text{out}}$, 
the wind has a constant local mass-loss rate. 
This gives a density structure at the base of the wind of
\begin{equation}
\nH^0 = \frac{0.5 \dot{M}_{\text{w}}}{2\pi 1.4 m_{\rm{H}} V_j R_0 \left( R_{\text{out}} -R_{\text{in}} \right)},
\label{density-base1}
\end{equation}
where $\dot{M}_{\text{w}}$ is the (two-sided) mass-loss rate of the wind. Note that for winds 
launched from a narrow region of the disk ($R_{\text{out}}  \simeq$ few $R_{\text{in}}$), 
the choice of the power-low index of the local mass-loss rate has a weak influence on 
$\nH^0$. In this work, we follow the modeling results of the SiO jet in HH212 \citep{2017A&A...607L...6T} and we fix $R_{\text{in}}=0.05$~au and $R_{\text{out}} =0.3$~au, leading to a density at the base of the streamlines of
\begin{equation}
\begin{split}
\nH^0 = 2.5 \times 10^{10} & \frac{\dot{M}_{\text{w}}}{10^{-6}M_{\odot}~\rm{yr}^{-1}} \left( \frac{0.25~\text{au}}{R_{\rm out} - R_{\rm in}} \right)  \\ & \left( \frac{V_j}{100~\rm{km~s}^{-1}}  \right)^{-1} 
 \left( \frac{0.15~\text{au}}{R_0}  \right)  \rm{cm}^{-3}.
\label{eq:density-base}
\end{split}
\end{equation}
\SC{To further reduce the parameter space,} we follow the universal correlation between accretion and ejection observed from Class 0 to Class II jets and set the \SC{(two-sided)} wind mass-flux to $\dot{M}_{\text{w}} =  0.1 \dot{M}_{\text{acc}}$.

\SC{In the end,} the parameter space \SC{in the present study} is thus reduced to only three free parameters: $\dot{M}_{\text{acc}}$, $M_\star$, and $Q/Q_{\text{ref}}$. 
To investigate how the chemical content of \SC{a dust poor, laminar} jet evolves in time with the decline of accretion rate and \SC{the increase in stellar mass} 
we computed six sets of wind model summarized in Table \ref{table:wind-models}, namely
\begin{enumerate}
\item Four models of a Class 0 wind. At this stage, the young embedded source has not reached its final mass yet and we choose $M_* =0.1 M_{\odot}$. Accretion rate is varied from $\dot{M}_{\text{acc}} = 5 \times 10^{-5} M_{\odot}~\rm{yr}^{-1}$ to $5 \times 10^{-6} M_{\odot}~\rm{yr}^{-1}$ to model sources with various accretion luminosities.
\item A model of a Class I wind. At this stage, the protostar has accumulated most of its mass and we choose $M_* = 0.5 M_{\odot}$ with an accretion rate of $\dot{M}_{\text{acc}}$ = $10^{-6} M_{\odot}~\rm{yr}^{-1}$.
\item A model of a Class II wind with an accretion rate of $\dot{M}_{\text{acc}} = 10^{-7} M_{\odot}~\rm{yr}^{-1}$ and $M_* =0.5 M_{\odot}$. This accretion rate is representative for actively accreting TTauri star.
\end{enumerate}
To study the impact of surviving dust, the five wind models have been computed for dust fraction $Q/Q_{\rm ref}$ = $0,10^{-3},10^{-2}$, and~0.1.

The initial chemical abundances at the base of the streamline are computed
as in Section \ref{sec:singlepoint}, ie. assuming chemical equilibrium and no attenuation of the radiation field. Hence they depend only on the adopted $T_{\rm K}$ = 1000 K, $T_{\rm vis}$ = 4000 K, and the initial ratios $\nH^0/G_0^0$ and $\nH^0/W^0$ given in Table \ref{table:wind-models} for all models.

\begin{figure*}[h]
\centering
\includegraphics[width=.8\textwidth]{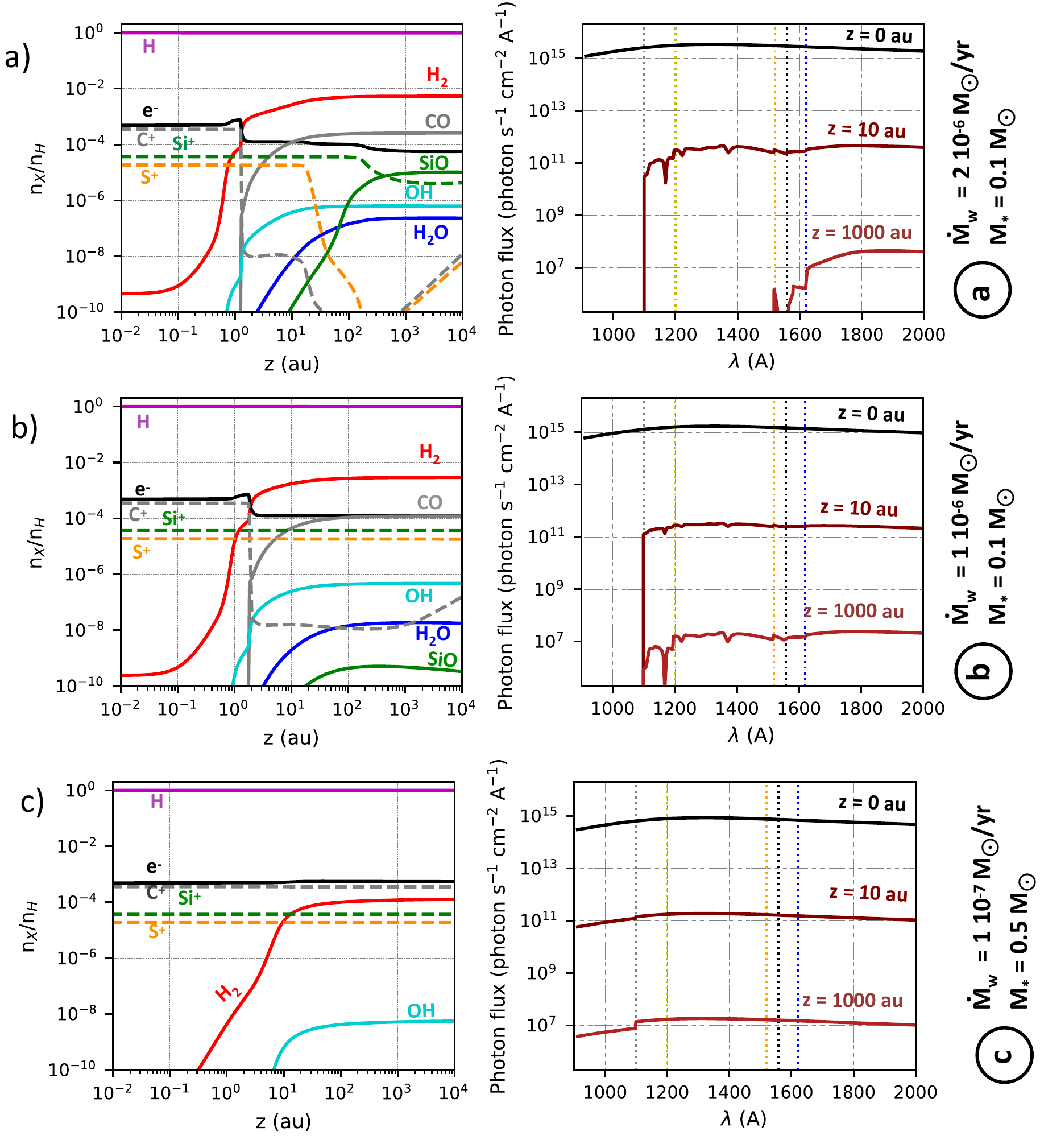}
\caption{Computed chemical abundances and local FUV radiation field for dust-free isothermal wind models for $T_{\rm{K}}=1000~$K. Left panels: chemical abundances relative to total H nuclei. Right panels: local mean intensity of the FUV radiation field at various position along the wind (position indicated on the curves). Photoionization thresholds of C, S, Si, Mg, and Fe are also indicated by gray, yellow, orange, black and blue dashed lines respectively. a) Class 0 model with  $\dot{M}_{\text{w}}$ = $2\times10^{-6}~M_{\odot}~\rm{yr}^{-1}$ and  $M_*$ = 0.1 $M_{\odot}$. b) Class 0 model with a lower mass-loss rate $\dot{M}_{\text{w}} = 10^{-6}~M_{\odot}~\rm{yr}^{-1}$ and same mass. c) Class I model, with lower accretion rate $\dot{M}_{\text{w}}$ = 10$^{-7}~M_{\odot}~\rm{yr}^{-1}$  but higher mass $M_* = 0.5  M_{\odot}$.}
\label{fig:wind-models-abund}
\end{figure*}

\subsection{Results: dust-free winds}
\label{subsec:wind-standard}

In this section, we present results on dust-free wind models ($Q=0$). Chemical abundances and local FUV radiation field of the selected wind models \textcircled{a}, \textcircled{b} and \textcircled{c} (see Table \ref{table:wind-models}) are presented in Fig. \ref{fig:wind-models-abund}. These specific models allow to highlight the influence of the wind parameters, namely the density of the wind (e.g mass-loss rate) and the radiation field. Models \textcircled{a} and \textcircled{b} have a different $\nH$ and $G_0$ but share the same $\nH/G_0$ ratio. Models \textcircled{b} and \textcircled{c} have a similar FUV radiation field but a different density (Table \ref{table:wind-models}).
Asymptotic abundances for the full set of dust-free wind models are also given in Fig. \ref{fig:wind-models-summary} (solid line).

\subsubsection{FUV field}

One of the main difference between single point models and wind models is the inclusion of the attenuation of the radiation field along streamlines. In dust-free winds, only gas-phase species can shield the radiation field and decrease photodissociation rates of molecular species.

Figure \ref{fig:wind-models-abund} (right panels) shows that the attenuation of the radiation field results in sharp absorption patterns characterized by thresholds below which the radiation field is heavily extincted. Those thresholds correspond to ionization thresholds of the most abundant atomic species. This specific attenuation pattern has already been pointed out by \citet{1989ApJ...336L..29G,1991ApJ...373..254G} in the context of stellar winds, although without focusing on the sharpness of the attenuation patterns that is at the root of the chemical richness of dust-free jets. The specific shielding mechanism that causes these unique attenuation patterns can be understood by the inspection the model \textcircled{b} (Fig. \ref{fig:wind-models-abund}-b).
In the absence of attenuation, carbon is expected to be ionized. However, at $z \simeq$ 2~au, the abundance of C$^+$ drops by several orders of magnitude (right panel) and neutral carbon (not shown here) becomes the main carbon carrier.
Figure \ref{fig:wind-models-abund}-b, right panel, shows that above this transition, photons below the photoionization threshold of carbon ($\lambda \le 1100$~\AA) are attenuated by more than 8 orders of magnitude. In this region of the spectrum, the opacity of the gas is dominated by carbon. 
The steep decrease of C$^+$ is thus due to the attenuation by carbon itself, triggering a C$^+$/C transition when the gas becomes optically thick to ionizing photons. Because of the increase of C at the C$^+$/C transition, the local opacity of the gas increases even more. This leads to an attenuation of the FUV field below $\lambda=$1100~\AA~that is much stiffer than attenuation by dust. This process, called continuum self-shielding and included in most of the dusty PDR models \citep{2007A&A...467..187R}, turns to be of paramount importance in dust-free and dust-poor winds.

The resulting attenuation of the FUV radiation field is very sensitive to the wind model. For model \textcircled{a}, with a higher mass-loss rate, the radiation field at $z=1000$~au is strongly attenuated down to the ionization wavelength threshold of sulfur, which has a relatively large value ($\lambda = 1600$~\AA), whereas for model \textcircled{c}, with a lower mass-loss rate, the radiation field is barely extincted across the FUV spectrum. As any self-shielding process, it depends on the column density of the \BG{neutral atom X. In the inner ionized and unattenuated part of the jet, this column density increases with z as
\begin{equation}
N_{\rm X}(z) = n_{\rm X}^0 z_0 \int_0^{z/z_0} \frac{(1+u)^2 + (\frac{z_0}{R_0} u)^2}{(1+u)^4} du
\label{eq:self-shielding-bg}
\end{equation}
where $n_{\rm X}^0$ is the density of X at the launching point and is proportional to $(\nH^0)^2/G_0^0$. The self shielding occurs if $N_{\rm X}(z)$ becomes typically larger than $\sigma_{\rm X}^{-1}$ where $\sigma_{\rm X}$ is the FUV absorption cross section\footnote{The FUV absorption cross sections of C, S, and Si are of the order of a few $10^{-17}$ cm$^2$.} of atom X. Because the integral term on the right hand side converges, eq. \ref{eq:self-shielding-bg} reveals a threshold effect: if $(\nH^0)^2/G_0^0$ is too low, $N_{\rm X}(z)$ is found to never rise above the critical value required to trigger the self-shielding, regardless of $z$.}

Models \textcircled{a} and \textcircled{b} share the same unshielded $\nH^0/G_0^0$ ratio and consequently exhibit similar atomic abundances at the base of the wind. However, model \textcircled{a} being denser, \BG{its $(\nH^0)^2/G_0^0$ ratio is larger. This increases} the total column density of S and Si \BG{which} exhibit self-shielding transitions. This results in attenuation of the radiation field at much longer wavelengths. In contrast, model \textcircled{c} having both a lower $\nH^0/G_0^0$ ratio and a lower density, the column density of carbon is not sufficient to attenuate the radiation field and the shape of the radiation field remains unaltered.

In other words, in the absence of dust, the radiation field is attenuated by continuum self-shielding of atoms. This process is very efficient at attenuating the radiation field below the photoionization thresholds of the atomic species. The wavelength below which the radiation field is attenuated depends on the species that are self-shielded. Carbon can attenuate radiation field at short wavelength whereas S, Si attenuate the radiation field at longer wavelength. Self-shielding by a specific species is very sensitive to the density of the wind. It results that Class II and I models are not dense enough to be shielded by any atom whereas Class 0 models are shielded by carbon for $\dot{M}_{\text{w}} \ge 5 \times 10^{-7} M_{\odot}~\rm{yr}^{-1}$ and by carbon, silicon and sulfur for $\dot{M}_{\text{w}} \ge 2 \times 10^{-6} M_{\odot}~\rm{yr}^{-1}$.

\subsubsection{H$_2$}
\begin{figure}
\centering
\includegraphics[width=.49\textwidth]{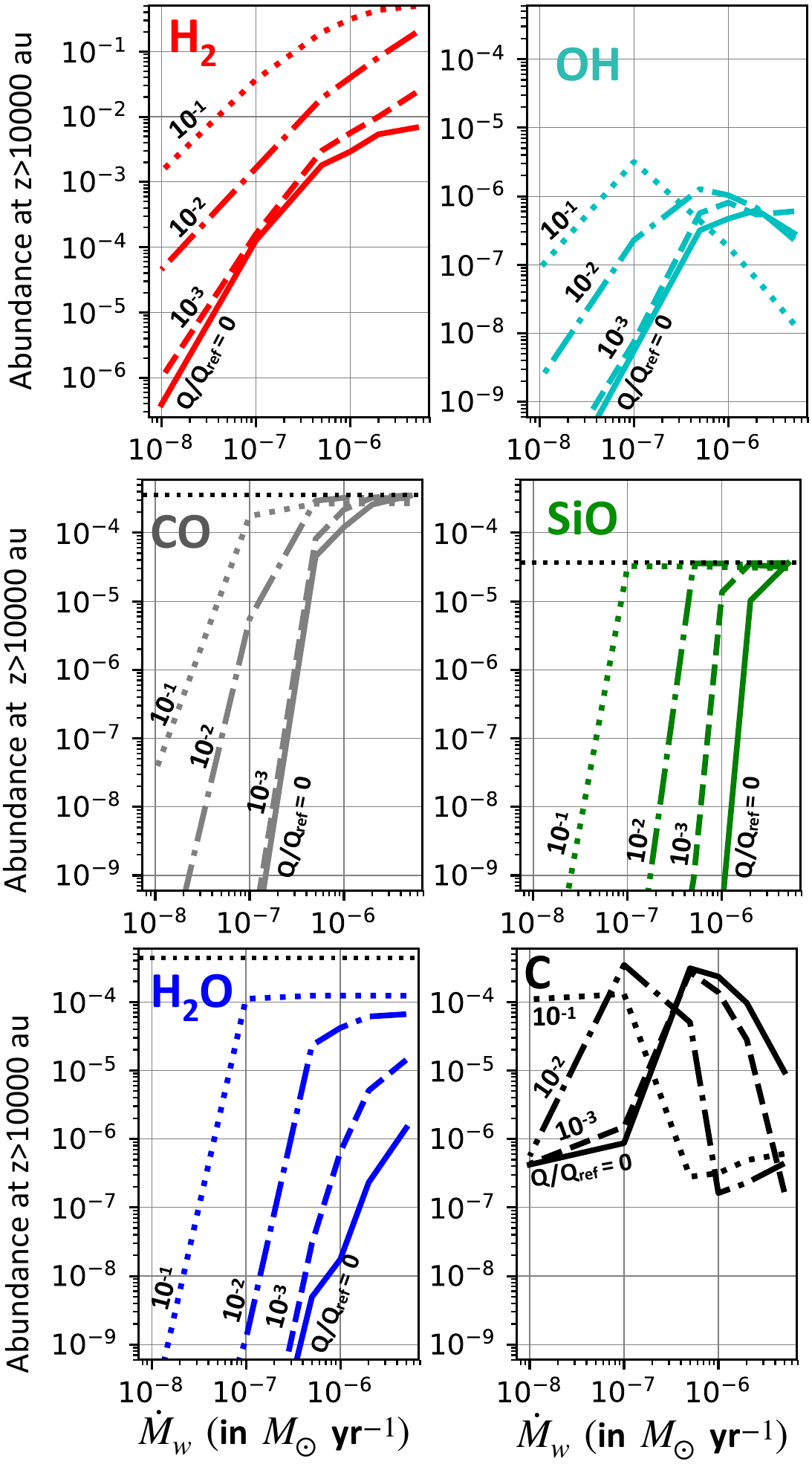}
\caption{Abundances at $z$ > 1000 au for a streamline anchored at 0.15 au in the disk as function of the mass-loss rate for various dust fraction. 
For $\dot{M}_{\text{w}} \ge 5 \times 10^{-7} M_{\odot}~\rm{yr}^{-1}$, the mass of the central object is $0.1 M_{\odot}$ (Class 0 model) and $0.5 M_{\odot}$ for lower mass-loss rates (Class I and II models). Dust-free models are plotted in solid lines, and dust-poor models with $Q/Q_{\text{ref}} = 10^{-3}, 10^{-2}$ and $0.1$ are plotted in dashed, dashed-dotted and dotted lines, respectively as indicated in each panel. Horizontal black dotted lines indicate the elemental abundance of carbon (panel on CO), silicon (panel on SiO), and oxygen (panel on H$_2$O).}
\label{fig:wind-models-summary}
\end{figure}

Figure \ref{fig:wind-models-abund} (left panels) and Fig. \ref{fig:wind-models-summary} show that dust-free wind models are poor in H$_2$ and consequently mostly atomic. H$_2$ is also smoothly increasing with increasing mass-loss rate. For Class 0 wind models, typical values of 10$^{-3}$ are found above $z=10$~au. This global trend with mass-loss rate is due to both an increase in the formation efficiency, and a decrease in destruction efficiency.

Regarding formation pathways, formation of H$_2$ is dominated by H$^-$ for all models, except at the base of the wind where H$_2$ is formed by CH$^+$ in Class I and II models. As seen in Section \ref{sec:singlepoint}, the efficiency of the H$^-$ route depends on the ratio $\nH/W$, that quantifies the ability of H$^-$ to form H$_2$ instead of being photodestroyed by the visible field. All models share the same visible radiation field so that the ratio $\nH/W$ depends only on the density of the wind (i.e. mass-loss rate). Consequently, the efficiency of the formation of H$_2$ by H$^-$ increases progressively with the mass-loss rate.

Along with the increase of the formation efficiency, the destruction efficiency decreases with mass-loss rate. For Class II models, destruction of H$_2$ are dominated by photodissociation. Class I models exhibit sufficiently high column density of H$_2$ to insure an efficient self-shielding, quenching photodissociation route. Alternatively, destruction by C$^+$ according to C$^+$ + H$_2 \rightarrow$ CH$^+ \rightarrow$ C$^+$ + H takes over from photodissociation with a smaller efficiency. For even larger mass-loss rates, the self-shielding of atomic carbon leads to a drop of C$^+$, quenching the former destruction route. Alternatively, H$_2$ is destroyed by atomic oxygen according to O + H$_2 \rightarrow$ OH $\rightarrow$ O + H.

In other words, dust-free wind models are mostly atomic. The H$_2$ abundance increases with mass-loss rate due to an increase of the efficiency of the formation route by H$^-$ and a decrease of the destruction route with the progressive self-shielding of H$_2$ and the recombination of  C$^+$.

\subsubsection{Other molecules}

Figure \ref{fig:wind-models-summary} shows that along with the increase of H$_2$ with the accretion rate, molecular abundances of interest increase with the mass-loss rate. As shown is Section \ref{sec:singlepoint}, H$_2$ abundance regulates the formation of OH, CO, SiO and H$_2$O.
On the other hand, continuum self-shielding of atomic species is a crucial process for the survival of these molecules: it reduces photodissociation rates by attenuating the FUV radiation field and quenches destruction routes by C$^+$. However, the precise impact of the shielding by atoms depends on the photodissociation threshold of each molecular species relative to the threshold below which the radiation field is attenuated.

OH is the first neutral species to be formed after H$_2$, and is an important intermediate for the synthesis of other molecules. Figure \ref{fig:wind-models-summary} shows that OH abundance is found to increase smoothly with mass-loss rate following the increase in H$_2$ abundance. Its abundance remains low, reaching $6 \times 10^{-7}$ for model \textcircled{a}. The wavelength dissociation threshold of OH being larger that the ionization thresholds of C, S, Si, Fe, and Mg, OH is not efficiently shielded by those species. Consequently, OH is destroyed by photodissociation at all $\dot{M}_{\text{w}}$ and the increase of OH with $\dot{M}_{\text{w}}$ is mostly driven by the smooth increase of H$_2$ and the increase of formation rates with density. Interestingly, at the highest mass-loss rate, the abundance of OH is also limited by the reverse reaction OH + H $\rightarrow$ O + H$_2$. 

CO is found to exhibit a much steeper increase with $\dot{M}_{\text{w}}$ from Class II-I models to Class 0 models (Fig. \ref{fig:wind-models-summary}). For Class 0 models, the CO abundance is high at $\ge 10^{-5}$. Interestingly, the abundance ratio CO/H$_2$ is $\simeq 2-5\times 10^{-2}$ in Class 0 models. This value is much larger than the canonical value derived in dusty molecular gas ($\simeq 10^{-4}$) and constitutes one the most striking characteristics of dust-free jets. The global behavior of CO is related to the shielding mechanism of the wind by carbon.
In contrast with OH, the CO wavelength dissociation threshold lays below the wavelength ionization threshold of carbon. CO can thus be efficiently shielded by carbon, when the column density of C is sufficient to trigger self-shielding of C. This process is notable in both Class 0 models presented in Fig. \ref{fig:wind-models-abund} where a jump in CO by several orders of magnitude is seen across the C/C$^+$ transition. Since self-shielding of carbon is only operating in Class 0, CO is abundant only in dust-free Class 0 models.

SiO exhibits an even steeper increase with mass-loss rate and is abundant only  for the highest mass-loss rates ($\dot{M}_{\text{w}} \ge 2 \times 10^{-7} M_{\odot}~\rm{yr}^{-1}$). Such a high sensitivity of SiO to the mass-loss rate is related to the self-shielding of silicon that ultimately controls formation and destruction of SiO. Regarding destruction, SiO has an intermediate wavelength dissociation thresholds ($\lambda_{th} = 1500~$\AA) that lies between C and Si ionization thresholds. As for OH, shielding by C does not reduce significantly photodissociation rates but in contrast to OH, shielding by S and Si is very efficient. As a consequence, the SiO photodissociation rate falls from model \textcircled{b} to model \textcircled{a}. Regarding formation routes, as shown in Sec. \ref{sec:singlepoint} and Fig. \ref{fig:molecule-formation-scheme}, SiO formation is more efficient when Si is in neutral form. Thus, the self-shielding of Si in model \textcircled{a} activates a direct and efficient formation route of SiO by Si + OH $\rightarrow$ SiO + H. On the contrary, in models with lower mass-loss rate, SiO synthesis proceeds via the much less efficient Si$^+$ route.

The H$_2$O abundance is found to be low, reaching $10^{-6}$ for the highest mass-loss rate. As for OH, the H$_2$O wavelength dissociation threshold is longer than the ionization threshold of C, S, Si and even Fe or Mg. H$_2$O is not efficiently shielded by the gas and its increase with mass-loss rate is mostly due to the increase of the H$_2$ abundance. At the highest mass-loss rates ($\dot{M}_{\text{w}} \ge 2 \times 10^{-6} M_{\odot}~\rm{yr}^{-1}$), H$_2$O is also destroyed though the reverse reaction H$_2$O + H $\rightarrow$ OH + H$_2$.

\subsection{Results: dust-poor winds}
\label{subsec:wind-dust-dust}

The inclusion of a non-vanishing fraction of dust activates H$_2$ formation on dust (see Section \ref{sec:singlepoint}), and introduces a new source of opacity for the radiation field.
Figure \ref{fig:wind-models-summary} summarizes the influence of dust fraction $Q/Q_{\text{ref}}$ on the asymptotic molecular abundances for the full set of wind models.

The inclusion of a small fraction of dust $Q/Q_{\text{ref}} = 10^{-3}$ increases H$_2$ abundances by a factor $\sim 2$ (Fig. \ref{fig:wind-models-summary}-a). Over the explored range of wind mass-loss rates, this specific dust fraction is indeed close to the critical value above which H$_2$ formation on grains takes over gas phase formation (see Sec. \ref{sec:singlepoint}). The chemistry of H$_2$ is consequently weakly affected. Similarly, the CO abundance is also weakly affected. This is because  the opacity of the gas below $\lambda \le 1100$~\AA~is still dominated by carbon over most of wind models. Consequently, dust does not affect CO photodissocitation rates and the CO abundance changes only by a factor $\simeq 3$. On the contrary, this small amount of dust has a strong impact on the abundance of SiO and H$_2$O.
Attenuation of the radiation field by this small amount of dust above $\lambda \ge 1100$~\AA~ shields SiO and H$_2$O, decreasing photodissociation rates. The attenuation by dust also 
triggers self-shielding of atomic species for lower-mass-loss rate ($\dot{M}_{\text{w}} = 10^{-6} M_{\odot}~\rm{yr}^{-1}$), leading to an even stronger attenuation of the radiation field at long-wavelengths. Furthermore, regarding SiO, the Si$^+$/Si transition activates the very efficient SiO formation route through Si. As a consequence, the critical mass-loss rate $\dot{M}_{\text{w}}$ above which the wind is rich is SiO is lowered. 

For larger $Q/Q_{\text{ref}}$ ratios, formation of H$_2$ on dust takes over from gas-phase formation.
Enhanced abundances of H$_2$ increase the formation rate of molecules whereas attenuation of the radiation field by dust reduces destruction rates. Figure \ref{fig:wind-models-summary} shows that CO, SiO, H$_2$O abundances increase by several orders of magnitude with increasing $Q/Q_{\text{ref}}$ above the critical value $Q/Q_{\text{ref}} = 10^{-3}$. 
The overall impact of dust on CO and SiO content is to lower the critical mass-loss rate below which the gas is rich in CO and SiO. For example, whereas dust-free wind models are rich in CO for $\dot{M}_{\text{w}} \ge 5 \times 10^{-7} M_{\odot}~\rm{yr}^{-1}$ and rich in SiO for $\dot{M}_{\text{w}} \ge 2 \times 10^{-6} M_{\odot}~\rm{yr}^{-1}$, dusty-wind models with $Q/Q_{\text{ref}} = 10^{-2}$ are rich in CO for $\dot{M}_{\text{w}} \ge 10^{-7} M_{\odot}~\rm{yr}^{-1}$ and rich in SiO for $\dot{M}_{\text{w}} \ge 5 \times 10^{-7} M_{\odot}~\rm{yr}^{-1}$. Interestingly, above those critical mass-loss rates, CO and SiO constitute the main carbon and silicon carriers. 

In contrast, the H$_2$O abundance has a more complex dependency on $Q/Q_{\text{ref}}$ and $\dot{M}_{\text{w}}$. H$_2$O reaches large abundances only for at high mass-loss rates and for rather large fraction of dust ($Q/Q_{\text{ref}} \ge 10^{-2}$). When the wind is sufficiently shielded by dust, destruction via the reverse reaction H$_2$O + H $\rightarrow$ H$_2$ + OH limits the abundance of H$_2$O. Being rich in atomic hydrogen, dust-free and dust-poor winds are hostile to the formation and survival of H$_2$O.

\subsection{Time dependent chemistry}

For all models, chemistry is found to be out-of-equilibrium, leading to asymptotic abundances that are somewhat smaller than steady-state abundances. As the density \BG{ and the radiation field drop with $z$ due to geometrical dilution and attenuation, the ratio between the chemical and the dynamical timescales increases as $V_j/(z\,\nH(z))$ for two-body reactions and as $V_j/(z\,F(z))$ for photoreactions, where $F(z)$ is the local FUV photon flux. The flow thus necessarily undergoes transitions beyond which part or all of the chemistry is "frozen". In the models presented here, we find that these transitions occurs around $z \simeq z_0$. We note, however,} that out-of-equilibrium effects do not explain the global trend with mass-loss rate (i.e. wind density) and with dust fraction, and only reduce the overall asymptotic abundances by a factor less than 4.

\subsection{Summary}

In this section, the chemistry of dust-free and dust-poor winds have been investigated by the use of parameterized wind models that include time-dependent chemistry and the attenuation of the radiation field. Our results \REV{for warm wind models ($T_{\text{K}} = 1000$~K)} are summarized in Fig. \ref{fig:wind-models-summary}. The overall molecular content of wind models increases with mass-loss rate of the wind and with the dust fraction. Dust-free and dust-poor winds are atomic but the small fraction of H$_2$, formed essentially via H$^-$ regulates the synthesis of other molecules. The survival of these molecules is insured by the attenuation of the radiation field by atomic species, and by dust, if any. The attenuation of the radiation field by atomic species proceeds through self-shielding, a process that depends critically on the column density of the absorbing species. It results in the presence of density or equivalently, mass-loss rate thresholds above which specific molecules are very abundant. Those thresholds depend on both the specific species and the dust fraction.

\section{Discussion}

\label{sec:disscussion}

Our results on disk wind chemistry have been obtained from a simple parametric wind model. The main advantage of this wind model is to be based on a simple geometry that allows a deep exploration of the parameter space and a detailed treatment of the radiative transfer. A number of limitations regarding the model presented here have to be taken into account before comparing our results to observations.

\subsection{Attenuation in the wind}

The radiative transfer is reduced to a 1D geometry by assuming that the radiation field is attenuated along the computed streamline. In a 2D geometry, the shielding of a streamline is provided by inner streamlines of the wind. This is especially true at the base of the wind, where photons coming from the accreting central object are impinging the streamline almost transversely. However, for most of the models presented here, the decline of the radiation field at the base of the wind $z < R_0$ is mostly caused by geometrical dilution, an effect that is properly taken into account by our 1D model. Species that contribute to the attenuation of the radiation field are mostly formed around $z \simeq 25 R_0$. At this distance, FUV photons propagate almost parallel to the streamlines and our 1D approximation is valid.
          
Our model also computes the attenuation of the radiation field by a limited number of species. As shown is Section \ref{sec:wind}, attenuation by atomic species is a key process for the survival of molecules. Because of their long-wavelength dissociation thresholds, molecules such as OH or H$_2$O can not be efficiently shielded by the considered atoms, namely C, Si, S, Mg, Fe. In this context, other elements that exhibit longer-wavelength dissociation thresholds such as Al, Ca, Li could shield important molecules. However, our models show that in the absence of dust the attenuation by Mg and Fe is negligible for $\dot{M}_{\text{w}} \le 10^{-6} M_{\odot}~\rm{yr}^{-1}$. Thus, rarer elements that have also \REV{similar photoionisation cross-sections} and similar recombination rates are expected to have a negligible contribution to the opacity of the gas.

\subsection{Thermal structure and shocks}
\label{subsec:discuss-shocks}

\SC{In this exploratory study,} we considered only isothermal \SC{laminar} disk wind models.  
The computation of thermal-balance is beyond the scope of this paper.
 \SC{The isothermal assumption is motivated by detailed thermal balance calculations in self-similar} MHD disk winds showing that ambipolar diffusion is a robust heating mechanism able to balance adiabatic \SC{and radiative} cooling, and naturally leads to a rather flat temperature profile \citep{1993ApJ...408..115S,2001A&A...377..589G}. Depending on the disk wind mass-loss rate, the asymptotic temperature \SC{for $R_0 \simeq 0.2$ au} varies from $500$~K to $4000$~K \citep{2012A&A...538A...2P,2016A&A...585A..74Y}, in line with our assumed fiducial temperature \SC{of 1000~K}. \SC{Similar calculations for a non-self-similar "X-wind" from the inner disk edge \citep{2002ApJ...564..853S} also show a quasi-isothermal behavior on inner streamlines; hence it appears to be a generic property of MHD disk winds that undergo large-scale collimation.} 
\SC{In contrast, models of stellar winds heated by ambipolar diffusion exhibited a decline of temperature at large distance
\citep{1990ApJ...361..546R}, due to the lack of magnetic collimation and the much smaller acceleration scale of a few $R_\star$ 
\citep[see discussion in Section 4.4.3 of][]{2001A&A...377..589G}.} 
Note that the shielding of the wind is not expected to be sensitive to the thermal structure of the jet, since it relies on electron recombination and charge exchange rate coefficients that do not strongly depend on the temperature. The local radiation field computed in this work \SC{thus} remains valid for warmer or cooler winds.
\SC{On the other hand, it should be kept in mind that asymptotic molecular abundances would be significantly lower at wind temperatures below 1000 K
(see Fig. \ref{fig:molec-chemistry}-a).}

\SC{In reality, shocks are likely to} also contribute to the heating \SC{and compression} of the gas, and impact 
molecule synthesis.
Millimeter and sub-millimeter interferometric observations of Class 0 molecular jets show that high velocity components of molecular lines are spatially resolved as a series of knotty structures. Detection of proper motions of these knots \citep{2015ApJ...805..186L}, as well as detailed kinematical properties \citep{2009A&A...495..169S,2017A&A...597A.119T} suggest that they are tracing internal shocks produced by time-variability \SC{in ejection velocity} \citep{1990ApJ...364..601R,1993ApJ...413..210S}. Numerical simulations including dust-free chemistry of H$_2$, but neglecting UV radiation field, shows that the H$_2$ abundance is indeed increased due to the increase of density in the post-shock gas \citep{2005RMxAA..41..137R}.

The effect of time-variability in the flow is not included in our simplified \SC{laminar} wind model. However, our results allow us to investigate \SC{qualitatively} the possible impact of internal shocks on molecule formation. In Sections \ref{sec:singlepoint} and \ref{sec:wind}, we show that high molecular abundances are reached in dust-free or dust-poor winds if it is warm with temperatures between $\simeq 800$ and $\simeq 3000$~K (see Fig. \ref{fig:molec-chemistry}-a), dense \REV{(see Fig. \ref{fig:molec-chemistry}-b)}, and well shielded by the gas \REV{(see Fig. \ref{fig:wind-models-abund})}. The impact of shocks is to increase locally the temperature and the density \REV{from their initial values}. Both effects drive up molecular abundances
compared to a laminar wind with the same \REV{equilibrium temperature and} average mass-flux. 
\REV{At larger distance}, all the successive internal shocks located between the protostar and $z$ would also increase the \SC{gas} attenuation of the \SC{radiation field}, increasing even more the molecular richness of the wind. \SC{However, on} large scale, the interstellar radiation field or the radiation field generated locally by strong shocks \citep[eg.][]{2012ApJ...751....9T} is expected to take over from the protostellar radiation field. \REV{Hence, asymptotic chemical abundances presented in Fig. \ref{fig:wind-models-summary}, neglecting the effect of shocks, should be considered as lower limits if the wind equilibrium temperature is $T_{\text{K}} \ge 800$~K, and if the protostellar radiation field dominates over other sources of FUV field. Otherwise, the peak abundances reached behind shocks could be lower than in Fig. \ref{fig:wind-models-summary} if molecule formation in the shocked gas is slower than the timescale to cool below $\simeq$ 800~K.} 

\REV{Therefore,} the precise quantitative impact of shocks remains to be constrained by detailed shock modeling, including self-consistent thermal balance. However, the overall decrease of molecular abundances with the decline of mass-loss rate and dust fraction remains a robust prediction of our model.

\subsection{Observational perspective}
\begin{figure}
\centering
\includegraphics[width=.47\textwidth]{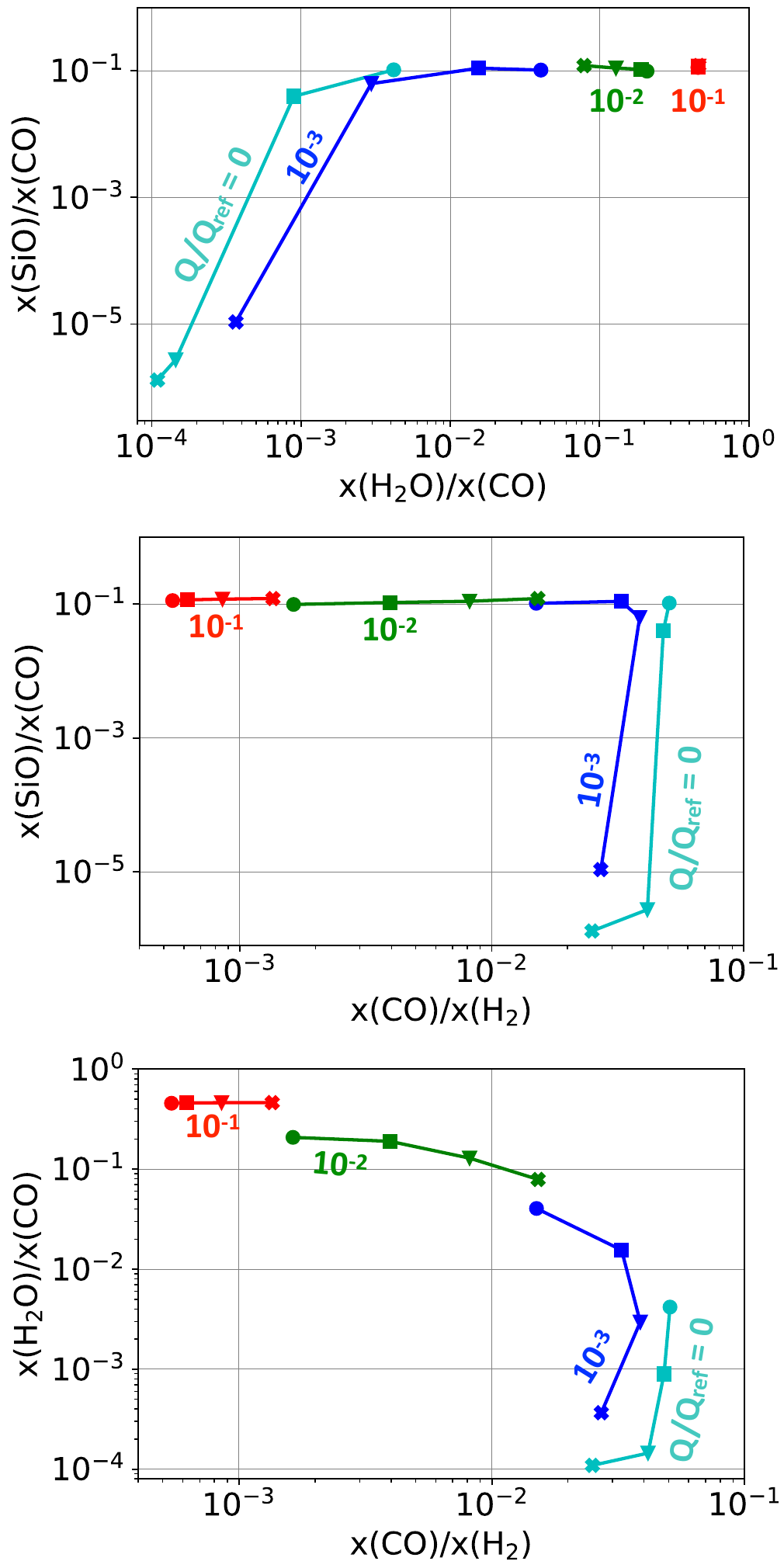}
\caption{\BG{Asymptotic abundance ratios for a streamline launched at $R_0=0.15$ au in the disk, for dust-free models (cyan) and dust-poor models with $Q/Q_{\text{ref}} = 10^{-3}$ (blue), $10^{-2}$ (green) and $0.1$ (red) and for different values of $\dot{M}_{\text{w}} = 5 \times 10^{-7}$ (crosses), $10^{-6}$ (triangles), $2 \times 10^{-6}$ (squares), and $5 \times 10^{-6}$ (circles) M$_{\odot}~\rm{yr}^{-1}$. All other parameters are kept constant to the values given in the footnote of Table \ref{table:wind-models}. Note that the variations of the ratios with $\dot{M}_{\text{w}}$ is due to the variations of $\nH^0$,
$\nH^0/G_0^0$, and $\nH^0/W^0$ with $\dot{M}_{\text{w}}$ as indicated in eqs. 
\ref{eq:density-base}, \ref{eq:G0-base}, and \ref{eq:W-base}.}}
\label{fig:abund-ratio}
\end{figure}

\begin{figure}
\centering
\includegraphics[width=.45\textwidth]{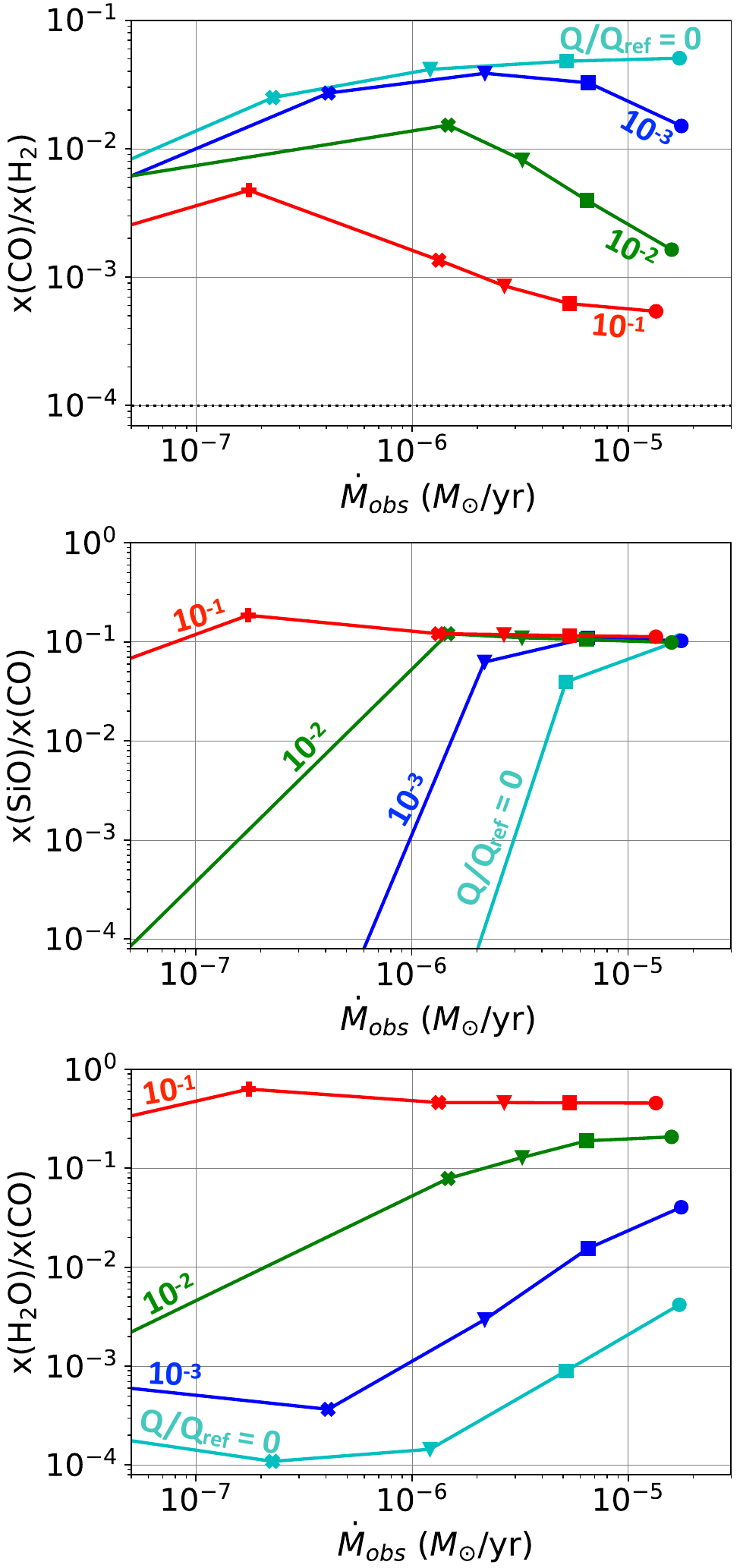}
\caption{\BG{Asymptotic abundance ratios for a streamline launched at $R_0=0.15$ au in the disk as function of the "observed" mass-flux $\Mco$ (see eq. \ref{eq:mco}), for dust-free models (cyan) and dust-poor models with $Q/Q_{\text{ref}} = 10^{-3}$ (blue), $10^{-2}$ (green) and $0.1$ (red) and for different values of $\dot{M}_{\text{w}} = 10^{-7}$ (straight crosses), $5 \times 10^{-7}$ (crosses), $10^{-6}$ (triangles), $2 \times 10^{-6}$ (squares), and $5 \times 10^{-6}$ (circles) M$_{\odot}~\rm{yr}^{-1}$. All other parameters are kept constant to the values given in the footnote of Table \ref{table:wind-models}. In particular $T_{\rm K}= 1000$ K.}}
\label{fig:abund-ratio-Mobs}
\end{figure}

\subsubsection{Mass-loss rates}

Clarifying the contribution of high velocity jets to mass and energy extraction during protostar formation remains an important observational challenge. Mass, momentum, and energy fluxes of deeply embedded Class 0 jets are usually derived from spectrally and spatially resolved CO rotational lines assuming a
\SC{standard reference CO} abundance of $x(\text{CO}) = x_\text{ref} \simeq 10^{-4}$ \citep[eg.][]{2016A&A...593L...4P}. 
\SC{Therefore, the observable quantity is not
the true wind mass flux $\Mw$ but the "observed" mass-flux $\Mco$ derived from CO observations assuming a standard CO abundance, namely
\begin{equation}
\Mco = \frac{x(\text{CO})}{10^{-4}} \Mw.
\label{eq:mco}
\end{equation}
}

Our models show that if the jet is dust-free or dust-poor, this assumption on the CO abundance \SC{may} lead to large errors on the derived energetic properties of jets. Indeed,  the mass\SC{-flux} will be overestimated by a factor $\sim~3.6$ \SC{when all of elemental carbon is in the form of CO} (i.e. for the highest $\dot{M}_{\text{w}}$ or $Q/Q_{\text{ref}} \simeq 0.1$), but underestimated by orders of magnitude at lower $\dot{M}_{\text{w}}$ or lower $Q/Q_{\text{ref}}$. 
Interestingly, we find that when the CO abundance is $x(\text{CO}) \ge 10^{-6}$, neutral carbon and CO are the main carbon carriers (see Fig. \ref{fig:wind-models-summary}). It suggests that sub-millimeter CI lines could be used together with CO to \SC{constrain the true CO abundance vs H and}
measure the true mass-flux of the jet. 

\SC{Alternatively, our models suggest that ratios of molecular abundances could be used in conjunction with the observed $\Mco$ 
to estimate the dust content of the wind and its true mass-flux. This is shown in the next section.}

\subsubsection{Dust content}
Our model results have shown that for a given wind temperature $T_{\text{K}}$, the molecular content is very sensitive to the precise dust fraction, suggesting that chemistry could be used as a unique tool to constrain the presence of dust in protostellar jets, and by way of consequence constrain their launching radius. Here we illustrate this point using our grid our laminar, isothermal disk wind models with a temperature of $T_{\text{K}} = 1000$~K, recalling that our exact results depend on specific choices of fixed
parameters given in the footnote of Table \ref{table:wind-models}.

\BG{We first display in Fig. \ref{fig:abund-ratio} the correlation between abundance ratios 
\SC{among brightest molecular jet tracers, namely CO/H$_2$, SiO/CO, H$_2$O/CO,} 
obtained for $0 \leqslant Q/Q_{\text{ref}} \leqslant 10^{-1}$
and $5 \times 10^{-7} \leqslant \dot{M}_{\text{w}} \leqslant 5 \times 10^{-6}$, i.e. for objects with high mass loss rates.}
\BG{For all dust poor models}, the $x(\text{CO})/x(\text{H}_2)$ ratio is found to be larger than $5 \times 10^{-4}$, 
in contrast with the standard ratio of $\simeq 10^{-4}$ derived in dusty molecular environments.
Such a high ratio would thus directly point toward a small dust fraction $\ll 1$. 
\SC{However, it may also be seen that all our models follow almost the same well-defined "path" in these ratio-ratio plots; in other words,
there is a strong degeneracy between $\Mw$ (in effect $n_H^0$) and $Q/Q_{\text{ref}}$
in the sense that the same ratio-ratio pair may be obtained by increasing one parameter and decreasing the other.
Such plots are therefore not able by themselves to determine the dust fraction $Q/Q_{\text{ref}}$ and true mass-flux
precisely. However, they do provide a good validation test of the underlying hypotheses in 
our simple laminar wind models, as all observations should lie close to the predicted curves.} 

\SC{In Fig. \ref{fig:abund-ratio-Mobs}, we present a more powerful observational diagnostic of the dust fraction, where the same ratios are plotted 
as a function of the "observable" (two-sided) mass-flux $\Mco$ (see eq. \ref{eq:mco}).
The plots start at $\Mco = 10^{-7}$ \Msun / yr, which is the minimum detectable mass-flux in CO assuming a threshold 
beam-averaged column density of $N_{\rm CO}= 3 \times 10^{15}$ cm$^2$, a beam size of $3 \times 10^{14}$ cm = 20~au, and a jet speed of 100 \kms.
It may be seen that the degeneracy between density and dust fraction is lifted, in different ways for the different ratios.}

The curve of $x(\text{CO})/x(\text{H}_2)$ vs $\Mco$ is found to provide a good diagnostic of the dust fraction, except for small $\Mco$ near the detection limit
or for very dust-poor winds with $Q/Q_{\text{ref}} \le 10^{-3}$ where curves come too close to discriminate.
Measuring $N(\text{CO})/N(\text{H}_2)$ in prototellar jets is challenging since pure rotational and ro-vibrational lines of H$_2$, and sub-millimeter lines of CO have very different upper energy levels and so are not necessarily tracing the same region of the jet. Probing simultaneously pure rotational lines of H$_2$ and high-J lines of CO would help to circumvent this problem. In that perspective, future JWST observations, in synergy with \textit{Herschel} data would help to probe the dust content of high velocity jets.

In contrast to CO/H$_2$, the SiO/CO ratio is very sensitive to the dust fraction \SC{even for small $\Mco$ and small $Q/Q_{\text{ref}}$}.
SiO is routinely observed though its submillimeter rotational transitions along with low-lying rotational transitions of CO \citep{1991A&A...243L..21B,1992A&A...265L..49G,2019arXiv191007857T}. These lines having similar upper energy levels, they can be used to derive abundance ratio, though opacity effects can complicate the analysis \citep{2012A&A...548L...2C}. 
\BG{Because of the increase of $\nH^0$ with $\dot{M}_{\text{w}}$ (see Eq. \ref{eq:density-base})}, our 1D laminar wind model predicted that SiO 
\SC{will reach its maximum abundance} 
only above a critical mass-loss rate that depends on the precise dust fraction \SC{(Fig. \ref{fig:wind-models-summary})}.
\SC{Figure \ref{fig:abund-ratio-Mobs} shows that even though the rise of SiO/CO with $\Mco$ is not as steep as $x$(SiO) versus $\Mw$ in Figure 10,
it would be possible to constrain the dust fraction from these curves if SiO/CO $< 0.1$.}
\BG{However, when the maximum possible amount of gas-phase SiO is reached for a given dust fraction, 
the SiO/CO abundance ratio is found to be remarkably constant with a value of about 0.1.}
\SC{In that case, the observed $\Mco$ only sets a lower limit to the dust fraction. 
In HH212, \citet{2012A&A...548L...2C} estimated a ratio SiO/CO \REV{$>0.04$}
while \citet{2007ApJ...659..499L} estimated $\Mco$ $\simeq 2 \times 10^{-6}$ \Msun /yr. This suggests a lower limit $Q/Q_{\text{ref}} \geq 10^{-3}$.} 
\BG{Joint observations of the SiO/CO and CO/H$_2$ abundance ratios together with $\Mco$ could
be used to derive the combination of mass-loss rate and dust fraction in protostellar jets.} 

\SC{We recall that} if SiO is formed in shocks \REV{and if the wind equilibrium temperature is $T_{\text{K}} \ge 800$~K}, \SC{the above diagnostic curves for laminar wind models will overestimate the true} dust fraction. Since SiO is shielded by S and Si, we also expect to find a correlation between high column density of atomic sulfur and silicon, and high abundances of SiO. In that perspective, observations of the [SI] line at $25.2 \mu$m  by JWST in synergy with ALMA will be of interest.

\SC{The ratio H$_2$O/CO vs $\Mco$} appears to be another good diagnostic for the dust content of the jets. Figure \ref{fig:abund-ratio} shows that the H$_2$O/CO abundance ratio depends on both the mass-loss rate and the dust fraction, and we predict rather low H$_2$O abundances in dust-poor jets. On-source water spectra have been acquired towards a large sample of low-mass protostars in the course of the WISH and WILL Key Programs \citep{2011PASP..123..138V,2017A&A...600A..99M}. HIFI observation have unveiled high-velocity bullets towards 4 out of 29 Class 0 sources of the WISH sample \citep{2012A&A...542A...8K}. The H$_2$O/CO abundance ratio is estimated to be $\simeq 1-10^{-1}$ for mass-loss rates of about $\Mco \simeq 10^{-5} M_{\odot}~\rm{yr}^{-1}$ \citep{2011A&A...531L...1K, 2010ApJ...717...58H}. \SC{Figure \ref{fig:abund-ratio-Mobs} shows that this is} in line with our laminar wind models with $Q/Q_{\text{ref}} \ge 3 \times 10^{-3}$.

Finally, since \SC{dust-poor} winds are shielded by S and Si below $\lambda = 1500$\AA, we predict that H$_2$O is photodissociated by longer-wavelength photons. Interestingly, it corresponds to the photodissociation of water through the A state that produces OH in vibrationally hot but rotationally cold state \citep{2001JChPh.114.9453V}. Thus, in the absence on an extra source of UV photons, OH would be expected to produce little mid-IR line from rotationally excited levels but strong ro-vibrational lines in the near-IR. This prediction can also be tested by future JWST MIRI and NIRSPEC observations. Again, shock models should be developed to properly derive the dust content in these molecular bullets. Comparison with other oxygen bearing species such as OH or atomic oxygen would also allow to better contain dust-poor models.

\subsection{Comparison with stellar wind models}

Our work demonstrates that disk winds launched within the silicate dust sublimation radius should be poor in H$_2$, \SC{with correspondingly high ratios of CO/H$_2$}. This unique feature of dust-poor chemistry was already identified in the context of stellar winds by previous authors \citep{1988MNRAS.230..695R,1989ApJ...336L..29G,1990ApJ...361..546R, 1991ApJ...373..254G}. \SC{With the exception of Ruden et al. (who had H$_2$ formation dominated by H$_2^+$ in their network) most of them also} found that photodetachement of H$^-$ is limiting the gas-phase formation rate of H$_2$.
However, our dust-free disk wind models predict H$_2$ abundances that are two orders of magnitude larger than those in dust-free stellar wind models with a similar mass-loss rate. 
For example, our Class 0 model (a) predicts an asymptotic H$_2$ abundance of $\simeq 5 \times 10^{-3}$ 
whereas the Case 1 model of \citet{1991ApJ...373..254G} predicts only $\simeq 2 \times 10^{-5}$. 
This difference is ultimately due to the geometry of the wind. 
For equal mass-loss rate and constant $V_w \simeq 100$ \kms, \SC{the initial $n_{\rm{H}}/W$ ratio in our disk wind model is similar to 
isotropic stellar winds  \citep[$\simeq 10^{13}$ cm$^{-3}$ for $\Mw \simeq 2-3 \times 10^{-6}$ \Msun /yr; see Case 1 in][]{1991ApJ...373..254G}. 
However, the asymptotic $n_{\rm{H}}/W$ ratio in the disk wind is increased by a large factor $(z_0/r_0)^2$ 
whereas it remains constant in the stellar winds.} This allows H$^-$ to survive and form H$_2$ in larger abundances.
On the other hand, a dust-free stellar wind that accelerates from the sonic point can reach an H$_2$ abundance closer to our model, 
as density at the wind base is higher for the same mass flux, leading to more efficient H$_2$ synthesis \citep[see Case 2 of][]{1991ApJ...373..254G}. 

Regarding the formation and survival of OH, CO, SiO and H$_2$O, most of the previous studies on dust-free stellar winds have discarded the impact of a UV excess, which turns to be of key importance in low-mass protostars \SC{with an accretion shock}. \SC{A notable exception is} \citet{1991ApJ...373..254G}, who have shown that the inclusion of a \SC{strong} FUV excess lead to a dramatic decrease of the abundances of H$_2$O, OH, SiO, whereas the abundance of CO was weakly affected. 
Our modeling \SC{explored for the first time a level of FUV excess adjusted self-consistently to the wind mass-flux, assuming a canonical ratio of 
$\Mw$ to $M_{\rm acc}$ of 0.1. It} shows that dust-free disk winds remain very rich in molecules above $\dot{M}_{\text{w}} \simeq 10^{-6} M_{\odot}~\rm{yr}^{-1}$
(for our adopted values of $T_{\text{K}}, R_{\rm in}$, $R_{\rm out}$, $V_j$, $R_\star$, etc). Even though an effect of the temperature is not excluded, our disk wind model also exhibits higher $n_{\rm{H}}/G_0$ ratio and higher total column densities (for similar $\dot{M}_{\text{w}}$) \SC{than stellar wind models}, causing \SC{more} efficient self-shielding by S and Si and \SC{favoring} the survival of molecular species against FUV photodissociation.

\section{Conclusion}

\label{sec:conclusion}

In this work, the chemistry and resulting molecular composition of disk winds launched within the dust sublimation radius has been explored by the use of single point and stationary wind models. 
Our results show that formation of H$_2$ through H$^-$ (electron catalysis) is a viable and dominant pathway to produce H$_2$ in dust-free disk winds. However, this route is not efficient enough to convert all atomic hydrogen into H$_2$. It results that dust-free winds are atomic, in line with stellar wind models \citep{1991ApJ...373..254G}. The inclusion of surviving dust has a significant chemical impact on H$_2$ for $Q/Q_{\text{ref}} \ge 10^{-3}$, where $Q_{\text{ref}} \equiv 6\times 10^{-3}$ is a reference dust-to-gas mass ratio. Above this value, the H$_2$ abundance in dust-poor winds depends on the dust fraction.
 
Our results also demonstrate that despite the low abundance of H$_2$, dust-free and dust-poor disk winds can be rich in molecules such as CO, SiO or H$_2$O. Molecular richness of dust-free winds depends critically on the properties of the wind. High temperatures ($T_{\rm{K}} > 800~$K) are required to activate efficient and direct formation routes of these molecules through OH. The overall molecular content of wind models increases with mass-loss rate of the wind and with the dust fraction. The survival of these molecules is insured by the attenuation of the radiation field by atomic species, and by dust, if any. The attenuation of the radiation field by atomic species proceeds through continuum self-shielding, a process that critically depends on the column density of the absorbing species. It results in the presence of density or equivalently, mass-loss rate thresholds above which specific molecules are very abundant. Those thresholds depend both on the specific species, and on the dust fraction. The order in which molecules are abundant as the density and/or the dust fraction increases is CO, SiO and H$_2$O. CO in abundant when continuum self-shielding of atomic carbon is triggered whereas SiO and H$_2$O are abundant when self-shielding of S and Si is activated.

Our results allow to propose observational diagnostics to probe the presence of dust in warm molecular jets ($T_{\text{K}} \ge 800~$K). As already suggested by \citet{1991ApJ...373..254G} in the context of stellar winds, a CO/H$_2$ abundance ratio larger that standard ISM value of $10^{-4}$ would be the primary evidence of dust-poor winds. The SiO abundance appears to be a \REV{promising} diagnostic to estimate the precise dust fraction in the wind: the SiO/CO abundance ratio exhibits a step increase from a critical $\dot{M}_{\text{w}}$ value that depends on the dust fraction. The H$_2$O/CO abundance ratio is also very sensitive to the dust fraction and to the density of the wind and could be used in combination with other oxygen bearing species.

The main limitation of \REV{our predictions} resides in the assumption of a laminar \REV{and isothermal} flow at $T_{\text{K}} =1000~K$. Indeed, observations at high angular resolution \REV{show} that molecules are present in shocks. Our models, together with \citet{2005RMxAA..41..137R}, suggest that internal shocks can locally enhance molecular abundances by increasing the temperature and the density of the gas. \REV{In the next paper of this series, shock models including the thermal balance will be investigated. Synthetic predictions of line intensities will allow to identify the most promising atomic and molecular transitions to constrain the presence of dust in jets. In that perspective, the present modeling constitutes a necessary step toward shock modeling by computing preshock conditions (chemical abundances and local UV field) from which shock models can be run.}  
  
Our work shows that chemistry is a powerful tool to unravel the dust content of protostellar jets. Future JWST observations, in combination with ALMA and \textit{Herschel} data will be able to provide key information on the launching region of the jets. If jets are dust-free or dust-poor, the chemical inventory of jets will give access to elemental abundances of the inner regions of protostellar disks and uncover key physical processes at stake during this early phase of star formation.

\begin{acknowledgements}
This work is part of the research programme Dutch Astrochemistry Network II with project number 614.001.751, which is (partly) financed by the Dutch Research Council (NWO). This work was also supported by the French program Physique et Chimie du Milieu Interstellaire (PCMI) funded by the Conseil National de la Recherche Scientifique (CNRS) and Centre National d'Etudes Spatiales (CNES).
\end{acknowledgements}

\bibliographystyle{aa} 
\bibliography{mybibli.bib} 

\begin{appendix}

\section{Radiation field}
\label{app:RF}

\begin{figure}[!h]
\centering
\includegraphics[width=.48\textwidth]{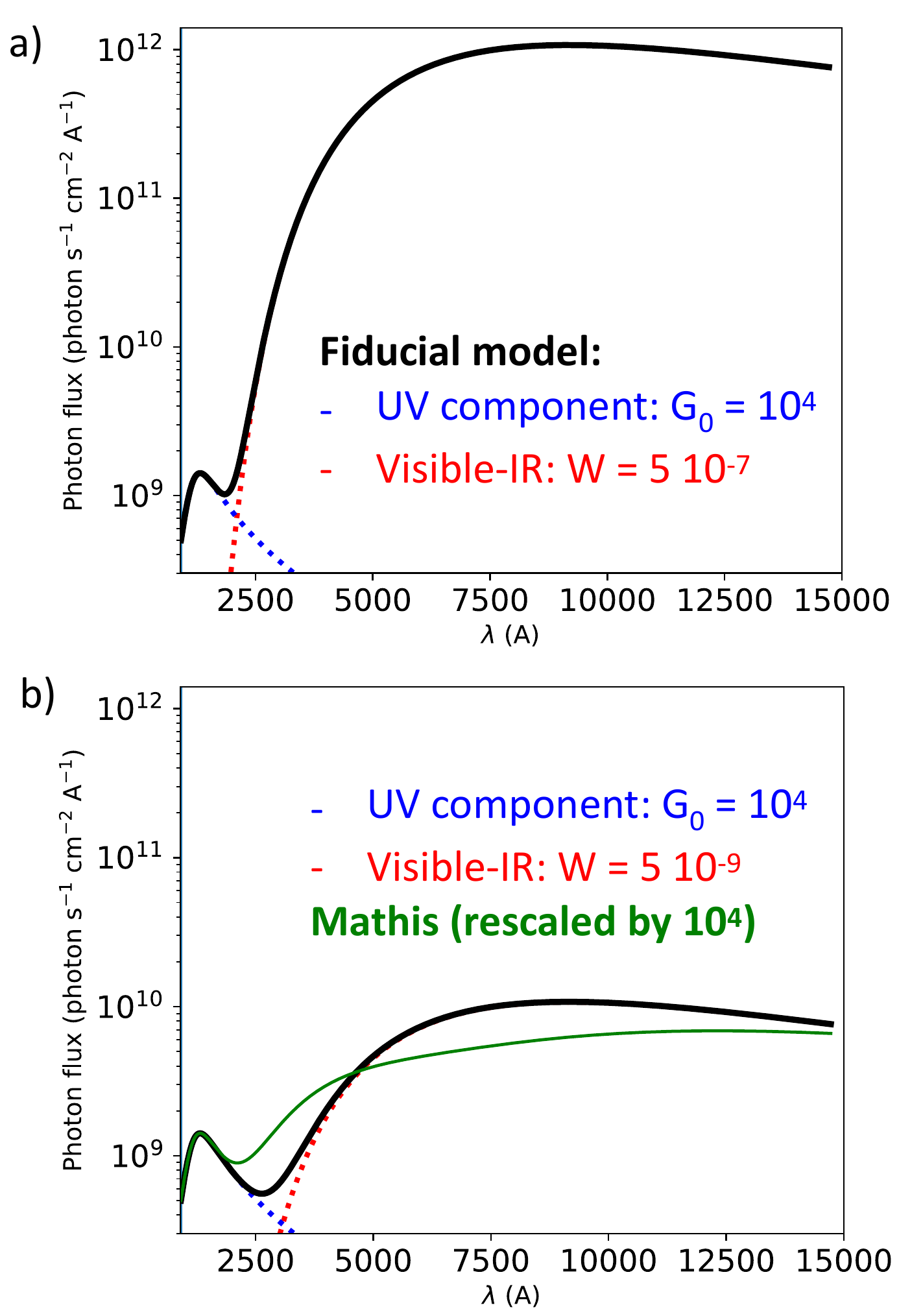}
\caption{Mean intensity of the adopted radiation field for two sets of parameter $\{ G_0,W \}$. The FUV and the visible components are in dotted blue and dotted red lines respectively and the total spectrum is in black solid line. a) Mean intensity of the radiation field adopted in Section \ref{sec:singlepoint} corresponding to $\{G_0=10^{4}, W = 5 \times 10^{-3}\}$. b) Mean intensity with the same FUV component but a weaker visible one ($\{G_0=10^{4}, W = 5 \times 10^{-9} \}$) overlaid on the Mathis ISRF scaled by $G_0 = 10^{4}$.
}
\label{fig:RF}
\end{figure}

Throughout this work, we model the radiation field by means of two components.
\begin{itemize}
\item A FUV component equal to the FUV part of the standard interstellar radiation field \citep[ISRF, ][]{1983A&A...128..212M} and rescaled by the parameter $G_0$. Following the fitting formula adopted in the Meudon PDR code\footnote{see \url{https://ism.obspm.fr}} \citep{2006ApJS..164..506L}, the FUV part is modelled by a mean specific intensity of
\begin{equation}
J_{\nu}^{FUV} = G_0 \times 107.2 \left( \text{tanh}  (4.07~10^{-3} \lambda - 4.6) + 1 \right)  ~\lambda^{-2.89},
\label{eq:ISRF-UV}
\end{equation}

with the wavelength $\lambda$ in \AA~and  $J_{\nu}^{ISRF, FUV}$ in $erg s^{-1} \rm{cm}^{-2}\AA^{-1}$. This formula fits the ISRF below $\lambda \simeq 2000$\AA~(see Fig. \ref{fig:RF}-b). 
\item  A visible-NIR component modeled by a black-body emission at a temperature $T_{\rm vis}=4000$~K, diluted by a factor W
\begin{equation}
J_{\nu}^{visible} = W \times B_{\nu}(T_{\rm vis}).
\end{equation}
In section \ref{sec:singlepoint}, $W$ is used as a parametrization of the visible-NIR intensity. In section \ref{sec:wind}, this parameter is the dilution factor of the photospheric emission and can thus be written as
\begin{equation}
W(R) = \frac{1}{2} \left( 1 - (1-R_*^2/R^2)^{1/2} \right),
\end{equation}  
where $R_*$ is the stellar radius and $R$ is the distance to the central object. For $R \gg R_*$ the dilution factor yields
\begin{equation}
W(R) = \frac{1}{4} \left( \frac{R_*}{R} \right)^2.
\end{equation}
\end{itemize}

Figure \ref{fig:RF}-a shows the radiation fields adopted in Section \ref{sec:singlepoint}.
Figure \ref{fig:RF}-b shows a radiation field with the same FUV component, but a weaker visible field  $\{G_0=10^{4}, W=5 \times 10^{-9}\}$ overlaid by the \citet{1983A&A...128..212M} ISRF scaled by $G_0=10^{4}$. The parametrized radiation field reproduces well the ISRF within a factor 2 from the visible ($\lambda \simeq 4000$~\AA) to the NIR ($\lambda \simeq 15000$~\AA). Thus, our parametrization gives also a good proxy for the shape of the ISRF for $W = 5\times10^{-13}$.


\section{Updated chemical network}
\label{app:chemical-network}

We review here chemical reactions and their adopted rate coefficients added to the chemical network of \citet{2019A&A...622A.100G}. Reaction rate coefficients that are outdated or inaccurate in standard astrochemical databases (KIDA, UMIST, OSU) have been fitted from theoretical and experimental work in the form of a modified Arrhenius law.

\begin{figure*}[!h]
\centering
\includegraphics[width=.32\textwidth]{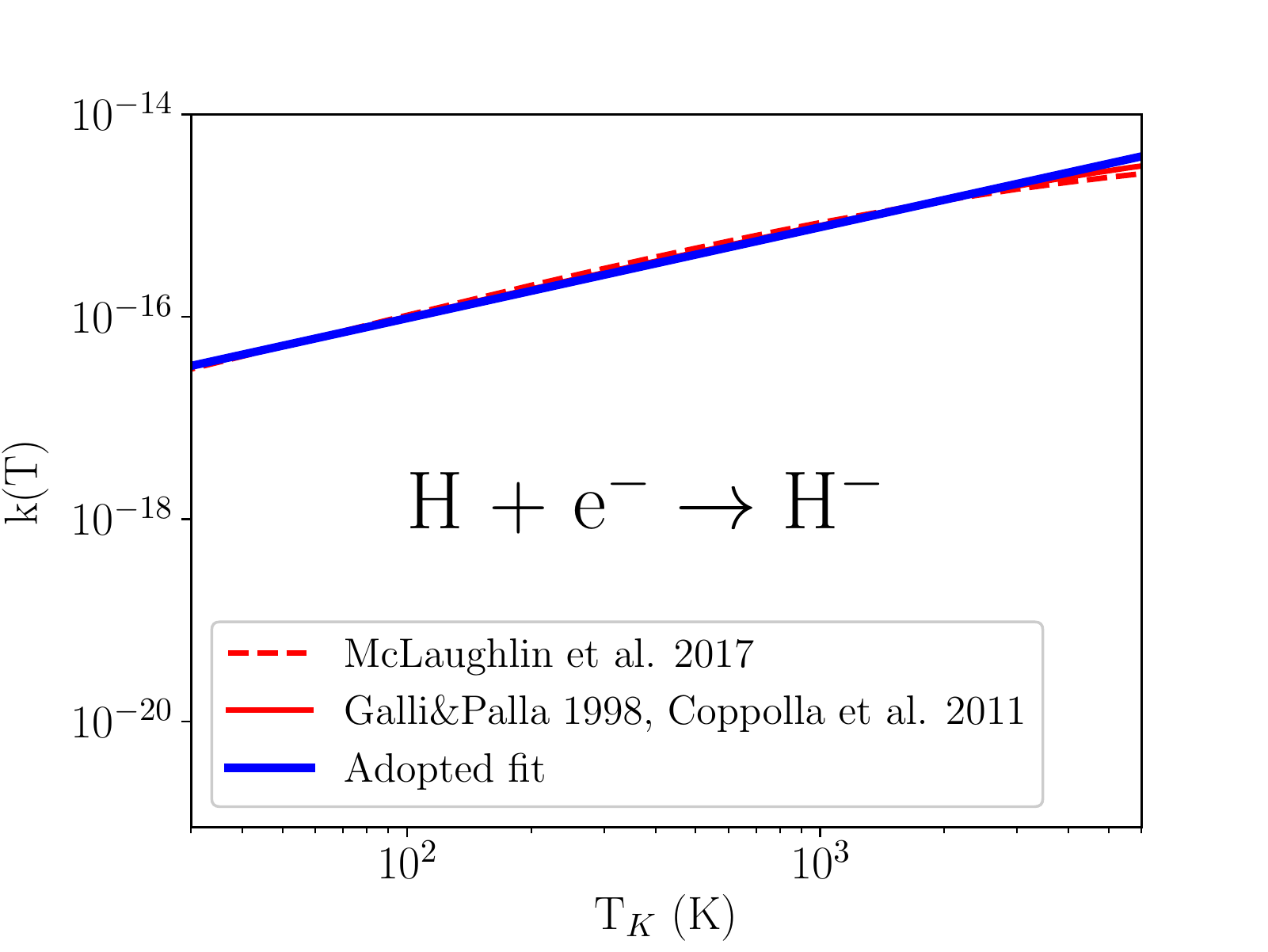} 
\includegraphics[width=.32\textwidth]{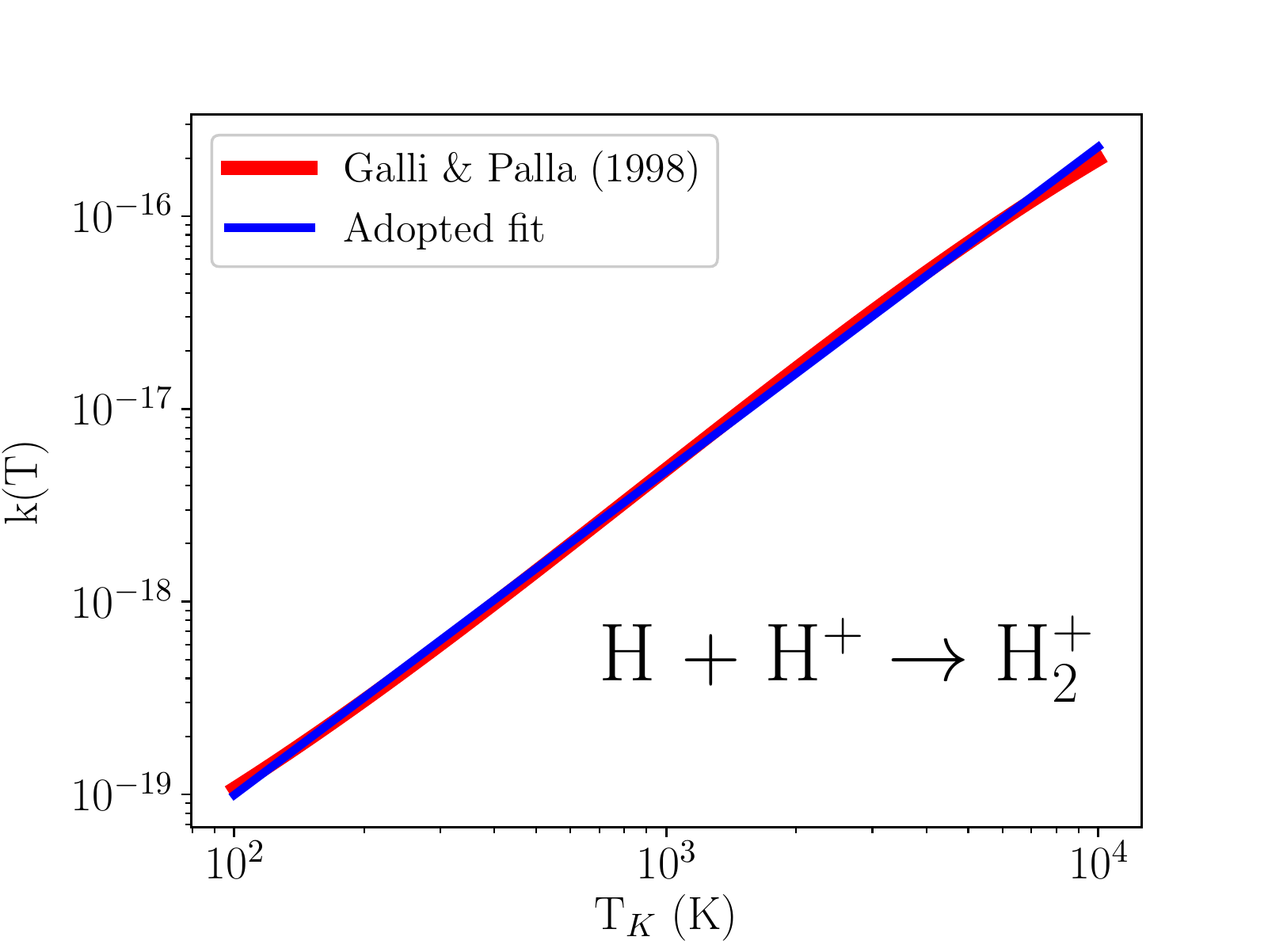} 
\includegraphics[width=.32\textwidth]{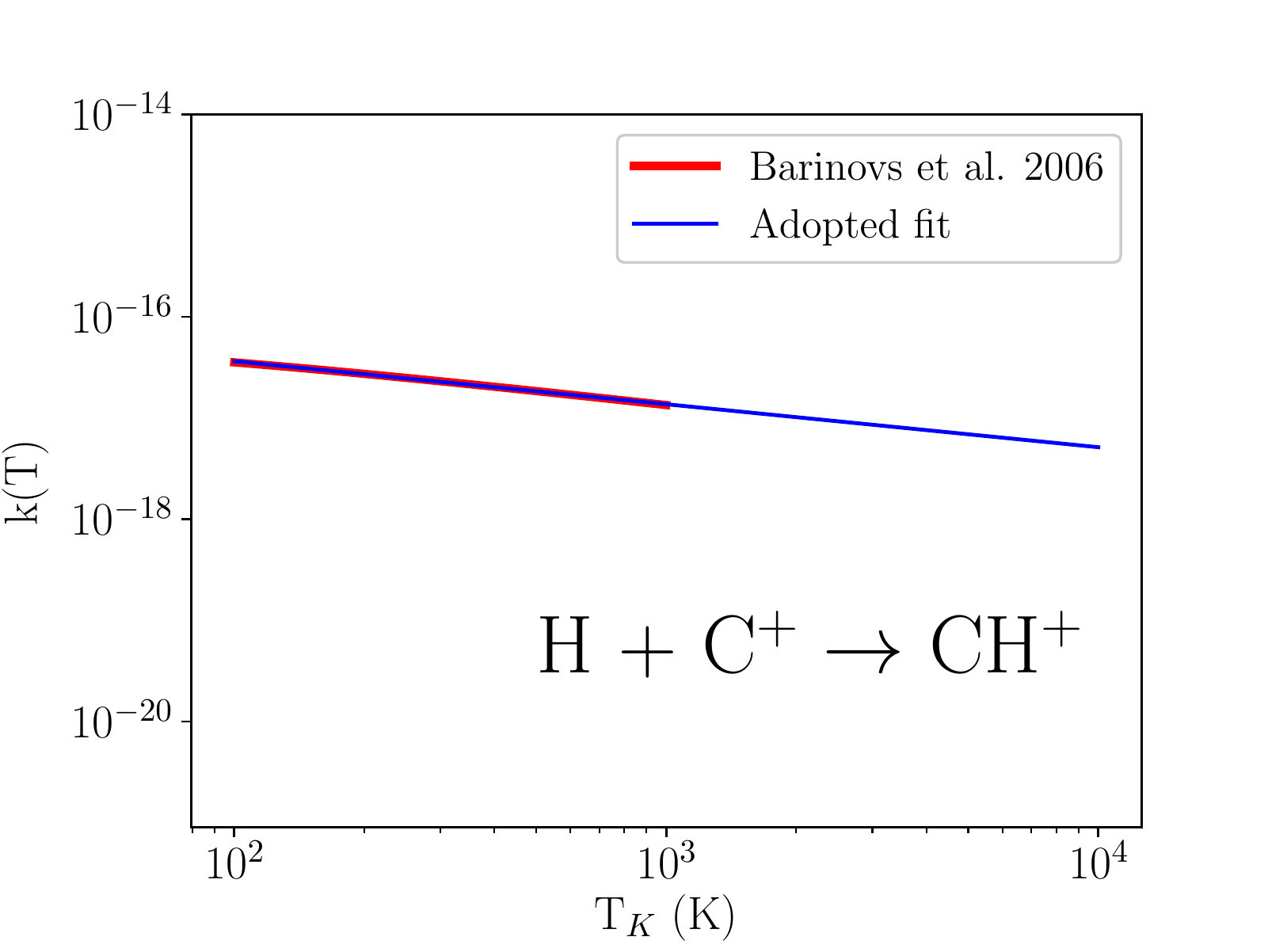} \\
\includegraphics[width=.32\textwidth]{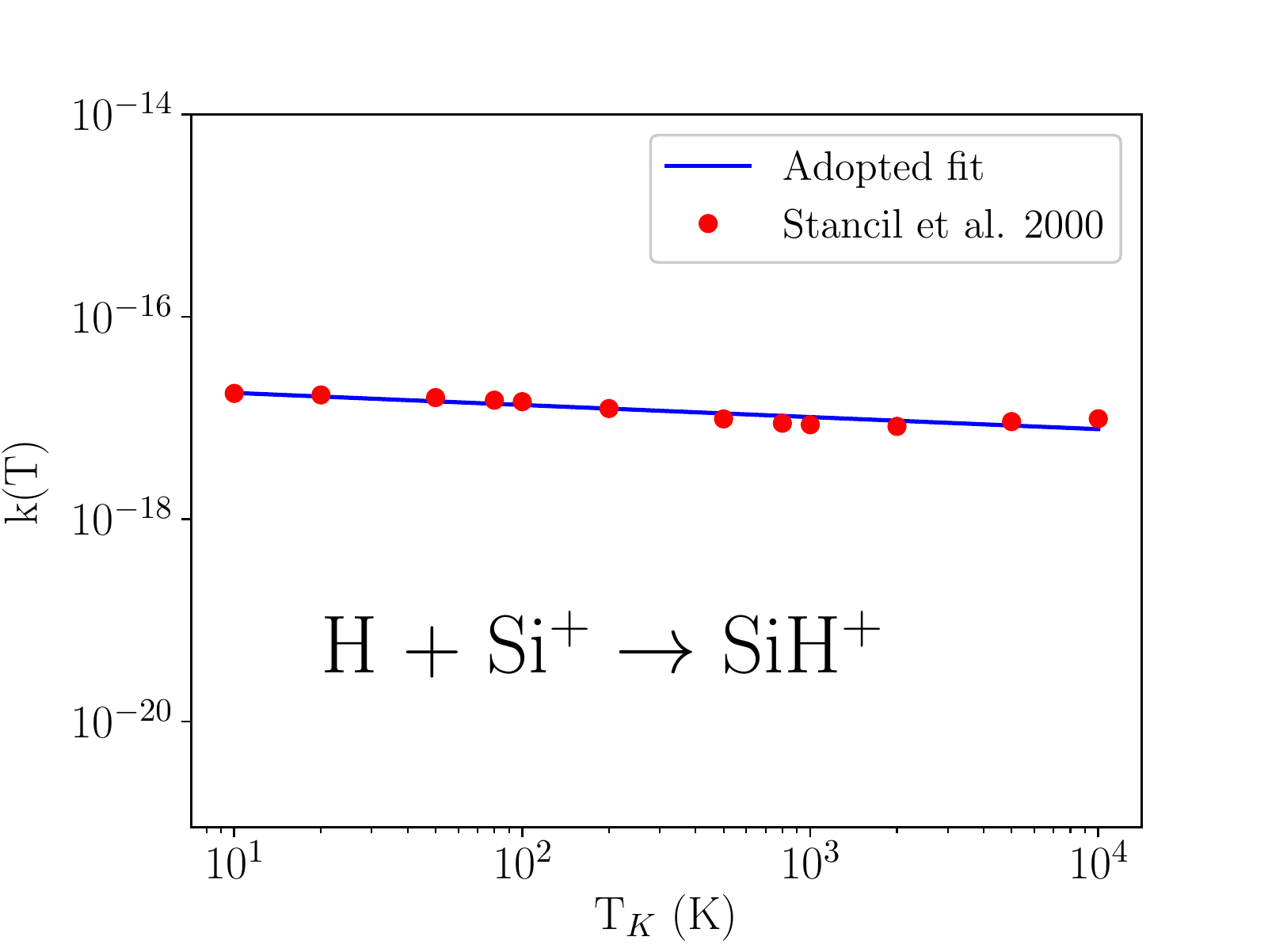} 
\includegraphics[width=.32\textwidth]{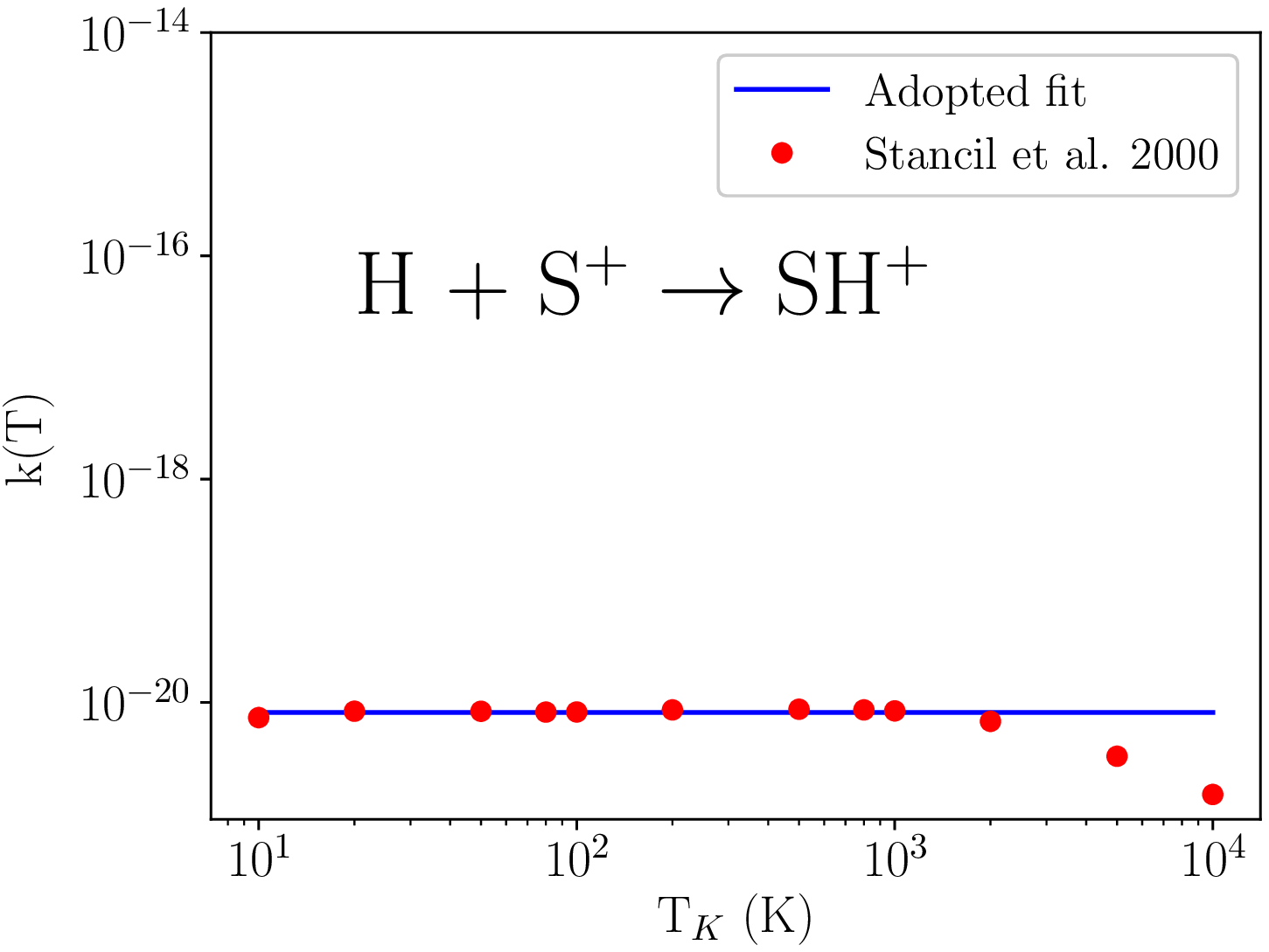}
\caption{Adopted radiative association rate coefficients. Red lines and red dots are computed or measured rate coefficients taken from references indicated in Table \ref{table:rad-asso}, and blue are the Arrhenius fits used in this work.}
\label{fig:rad-asso}
\end{figure*}

\paragraph{Radiative association.}
\label{app:chemical-network-rad-asso}

Gas-phase formation routes of H$_2$ involving two-body reactions are initiated by radiative associations. As shown in Appendix \ref{app:anal-H2}, the efficiency of these routes is directly proportional to the radiative association rate coefficient. It is thus crucial to get accurate rate coefficients, especially for the radiative association that is found to be dominant, namely between hydrogen and electron and between hydrogen and C$^+$. Table \ref{table:rad-asso} compiles the radiative association rate coefficients adopted in this work together with references used to fit the rate coefficients. Original data and adopted fits are plotted in Fig. \ref{fig:rad-asso}.

\begin{table}[!h]
\centering
\caption{Radiative association rate coefficients with hydrogen relevant for the formation of H$_2$.}              
\label{table:rad-asso} 
\begin{tabular}{c c c}    
\hline        
Reaction   & Rate ($\rm{cm}^3 \rm{s}^{-1}$)  & Ref.\\  
\hline                                  
e$^- + $H  &  2.6 $\times$ 10$^{-16}$ $\left( \frac{T_{\rm{K}}}{300} \right)^{0.9}$ &  (1) \\
C$^+ + $H  &  2.28 $\times$ 10$^{-17}$ $\left( \frac{T_{\rm{K}}}{300} \right)^{-0.42}$ & (2) \\
Si$^+ + $H &  $1.18 \times 10^{-17} \left( \frac{T_{\rm{K}}}{300} \right)^{-0.12}$ &  (3) \\
S$^+ + $H  &  $8.13  \times 10^{-21}$ &  (3) \\
H$^+ + $H  &   $6.3 \times 10^{-19} \left( \frac{T_{\rm{K}}}{300} \right)^{1.68}$ & (4,5) \\
\hline
\end{tabular}
\tablebib{
(1)~\citet{2017JPhB...50k4001M} ; (2)~\citet{2006ApJ...636..923B}; (3)~\citet{2000A&AS..142..107S}; (4)~\citet{1973AD......5..167R}; (5) \citet{1993ApJ...414..672S}
}
\end{table}

\paragraph{Other chemical reactions associated to H$_2$.}
Other reactions related to H$_2$ formation and added to the original network are reported in Table \ref{table:other-reaction-H2}.

\begin{table}
\caption{H$_2$ related chemical reactions added to \citet{2019A&A...622A.100G}. Photodetachement rate of H$^-$ is given for a 4000 K black-body radiation field diluted by a factor W.}             
\label{table:other-reaction-H2} 
\begin{tabular}{c c c}    
\hline        
Reaction & Rate (cm$^{3}$ s$^{-1}$) & Reference \\    
\hline                                   
H$^- + $H $\rightarrow$ H$_2 +$ e$^-$ &  $1.3(-9)$ & KIDA \\
H$^- + $h$\nu \rightarrow$ H $+$ e$^-$ &  $6.0(+5) W$s$^{-1}$ & (3) \\
H$^- + $H$^+$ $\rightarrow$ H $+$ H & $2.3(-7)(T_{\rm{K}}/300~\rm{K})^{-0.5}$ & KIDA \\
H$^- + $H $^+ \rightarrow$ H$_2^+ +$ e$^-$ &  $8.4(-9)(T_{\rm{K}}/300~\rm{K})^{-0.35}$  & (4) \\
\hline                                             
H$+$H$+$H $\rightarrow$ H$_2$ + H &  $2.6 \times 10^{-32} \left( \frac{T_{\rm{K}}}{300~\rm{K}}\right)^{-0.34}$ &  Fit from (5)\\
\end{tabular}
\tablebib{
(1)~\citet{1998JPhB...31.2571C}; (2) \citet{2011ApJS..193....7C}; (3) \citet{2017A&A...602A.105H};(4)\citet{1998A&A...335..403G} ;(5) \citet{2013ApJ...773L..25F}.
}
\end{table}

\paragraph{Mg charge-exchange network.}
Chemical reactions controlling the ionization state of magnesium added to the original chemical network are reported in table \ref{table:Mg}.

\begin{table*}
\centering
\caption{Magnesium chemical network added to \citet{2019A&A...622A.100G}. All reaction but photoionization are from KIDA database.}              
\label{table:models} 
\begin{tabular}{c c }    
\hline       
Reaction & Rate (cm$^3$ s$^{-1}$) \\    
\hline                                   
Mg $+$ h$\nu \rightarrow$ Mg$^+ +$ e$^-$ & Cross-section from (1) \\
\hline 
Mg$^+ +$ e$^-    \rightarrow$ Mg + h$\nu$ & $2.80\times 10^{-12} (T_{\rm{K}}/300 K)^{-0.86}$ \\
\hline 
Mg $+$ H$^+      \rightarrow$ Mg$^{+} +$ H &                  $1.1 \times 10^{-9}$ \\
Mg $+$ H$_3^+    \rightarrow$ Mg$^{+} +$ H$_2 +$ H  &         $1   \times 10^{-9}$ \\
Mg $+$ C$^+      \rightarrow$ Mg$^{+} +$ C &                  $1.1 \times 10^{-9}$ \\
Mg $+$ HCO$^+    \rightarrow$ Mg$^{+} +$ HCO &                $2.9 \times 10^{-9}$ \\
Mg $+$ H$_3$O$^+ \rightarrow$ Mg$^{+} +$ H$_2$O + H &         $1   \times 10^{-9}$ \\
Mg $+$ H$_2$O$^+ \rightarrow$ Mg$^{+} +$ H$_2$O &             $2.2 \times 10^{-9}$ \\
Mg $+$ O$_2^+    \rightarrow$ Mg$^{+} +$ O$_2$ &              $1.2 \times 10^{-9}$ \\
Mg $+$ NO$^+     \rightarrow$ Mg$^{+} +$ NO &                 $8.1 \times 10^{-10}$ \\
Mg $+$ S$^+      \rightarrow$ Mg$^{+} +$ S &                  $2.8 \times 10^{-10}$ \\
Mg $+$ SH$^+     \rightarrow$ Mg$^{+} +$ SH &                 $2.6 \times 10^{-9}$ \\
Mg $+$ SO$^+     \rightarrow$ Mg$^{+} +$ SO &                 $1   \times 10^{-10}$ \\
Mg $+$ H$_2$S$^+ \rightarrow$ Mg$^{+} +$ H$_2$S &             $2.8 \times 10^{-9}$  \\
\hline
Mg$^+ +$ H $     \rightarrow$ Mg$^{+} +$ H$^+ +$ e$^-$ &      $1.3 \times 10^{-13} (T_{\rm{K}}/300 K)^{-0.86}$ e$^{-157890/T}$ \\
Mg$^+ +$ H$_2    \rightarrow$ Mg$^{+} +$ H$_2^+ +$ e$^-$ &    $1.1 \times 10^{-13} (T_{\rm{K}}/300 K)^{-0.86}$ e$^{-179160/T}$   \\
Mg$^+ +$ H$_2    \rightarrow$ Mg$^{+} +$ H $+$ H $+$ e$^-$ &  $3   \times 10^{-11} (T_{\rm{K}}/300 K)^{-0.86}$  e$^{-52000/T}$  \\
Mg$^+ +$ He $    \rightarrow$ Mg$^{+} +$ He$^+ +$ e$^-$ &     $1.1 \times 10^{-13} (T_{\rm{K}}/300 K)^{-0.86}$  e$^{-285328/T}$  \\
\hline                                             
\end{tabular}
\tablebib{
(1) \citet{2017A&A...602A.105H}.
}
\label{table:Mg}
\end{table*}

\section{Analytical approach of H$_2$ formation in gas-phase}
\label{app:anal-H2}
In this appendix we propose an analytic approach for the formation of H$_2$ in dust-free and dust-poor environments. Our goal is to derive analytical expression of boundaries defining the dominant formation route in the $\{n_{\rm{H}}, W \}$ plane (see Fig. \ref{fig:H2-summary}), and derive the analytical expression of the abundance of H$_2$ plotted in Fig. \ref{fig:H2-chemistry}.
To do so, formation rates of H$_2$ by H$^-$ and by any XH$^+$ ion are first derived, taking into account photodestruction of these intermediates. Then, a comparison between various routes, including three-body reaction is given.

\subsection{Formation rate by gas phase catalysis}

Formation of H$_2$ though H$^-$ and any ion XH$^+$ (with X=C, S, H...) is a two stage catalytic process. In this section, we derive the resulting formation rate of H$_2$ assuming steady state for the intermediate species H$^-$ and XH$^+$ (Bodenstein approximation). Throughout this section, the gas is assumed to be neutral and atomic ($x(H) \simeq 1$), in line with our modeling results (Section \ref{sec:singlepoint}, \ref{sec:wind}).

\paragraph{Electron catalysis.} 

H$_2$ can be formed through the intermediate anion H$^-$, via a slow radiative attachment and a fast associative detachment:
\begin{align}
&\Hy + \text{e}^- \rightarrow \Hy^- + h \nu~~~~~~~~~~~~~~k_1^e(T)~,  \label{eq:rad-el-asso} \\
&\Hy^- + \Hy \rightarrow \Hy_2 + \text{e}^-~~~~~~~~~~~~~~k_2^e(T)~. \label{eq:asso-det}
\end{align}

The efficiency of this route depends on the survival of the intermediate H$^-$. In neutral medium, the main destruction route that can compete with the associative detachment (\ref{eq:asso-det}) is the photodetachment of the fragile H$^-$ anion by visible photons,
\begin{align}
\Hy^- + h\nu \rightarrow \Hy + \text{e}^-. \label{eq:photo-det}
\end{align}
For a diluted black-body at $4000$~K, H$^-$ decays preferentially through the associative detachment (\ref{eq:asso-det}) to produce H$_2$ if
\begin{align}
\nH/W > \nH^{crit} \equiv \frac{k_{\phi}^{0,e}}{k_2^e}  =  4.6 \times 10^{14}~\rm{cm}^{-3},
\label{eq:photo-det-condition}
\end{align}
where $k_{\phi}^{0,e}$ is the photodetachment rate by an undiluted black-body at $4000$~K.
Above this critical value, the efficiency of the electron catalysis is optimal. Defining
\begin{equation}
\eta \equiv \frac{\nH}{W n_{crit}^{W}},
\label{eq:eta-def}
\end{equation}
a fraction 
\begin{equation}
\frac{\eta}{1+\eta} 
\end{equation}
of the H$^-$ formed by radiative association is converted in H$_2$. The associative detachment or photodetachment being very fast, H$^-$ reaches steady-state abundances much faster than H$_2$, in a typical time scale of \footnote{At first order, stationary time scale of a chemical species corresponds to its destruction time scale.} 
\begin{align}
t_{\text{H}^-} \lse (k_1^e \nH)^{-1} \simeq 8 \left( \frac{\nH}{10^8 \rm{cm}^{-3}}  \right)^{-1} ~ \rm{s}.   
\end{align}

For $t \gg t_{\text{H}^-}$, the resulting H$_2$ formation rate of the full catalytic process, valid even when H$_2$ abundance is out-of-equilibrium is: 
\begin{align}
&R^{H_2}_{elec} =  3.4 \times 10^{-19} \left( \frac{x_e}{4.8~10^{-4}}  \right) \left( \frac{T_{\rm{K}}}{1000 K}\right)^{0.9}\frac{\eta}{1+\eta} \left(\frac{\nH}{1 \rm{cm}^{-3}} \right)^2 \rm{cm}^{-3} \rm{s}^{-1}.
\label{eq:formation-elec-cata}
\end{align}
The efficiency of this route is directly proportional to the electron fraction. For $\eta \gg 1$, this rate does not depend on the rate of the fast associative attachment (\ref{eq:asso-det}) but only on the rate of the limiting slow radiative association (\ref{eq:rad-el-asso}), leading the a rate of
\begin{align}
&R^{H_2}_{elec} =  3.4 \times 10^{-19} \left( \frac{x_e}{4.8~10^{-4}}  \right) \left( \frac{T_{\rm{K}}}{1000 \rm{K}}\right)^{0.9}   \left(\frac{\nH}{1 \rm{cm}^{-3}} \right)^2 ~~\rm{cm}^{-3} \rm{s}^{-1}.
\label{eq:formation-elec-cata-2}
\end{align}

\begin{table}
\caption{Formation rates of H$_2$ by electron and ionic catalysis in cm$^{-3}$ s$^{-1}$. Rates for electron and  ionic catalysis are parametrized according to equation  (\ref{eq:formation-elec-cata}) and (\ref{eq:ionic-catalysis}) respectively. $\nH^{crit, X}$ is defined in equation (\ref{eq:eta}). $\alpha$ is the H$_2$ formation rate in $\rm{cm}^{-3} \rm{s}^{-1}$ at $T_{\rm{K}}=1000$~K assuming a reference abundance of the catalyst of $x_X$. This reference value is the elemental abundance of the corresponding element for X=C, S, Si, and an abundance representative of dust-free wind for  H$^+$. The electron abundance corresponds to the one obtained if all atoms with an IP<13.6eV are ionized. Numbers in parentheses are powers of 10.} 
\label{table:H2-rates} 
\begin{tabular}{c c c c c}    
\hline        
Catalyst & x$_{X}$ & $\alpha$ ($\rm{cm}^{-3} \rm{s}^{-1}$) & $\beta$ & $\nH^{crit, X}$ (cm$^{-3}$) \\    
\hline                                   
e$^-$ & $4.8 (-4)$ & $3.35(-19)$ & $0.9$ & . \\
C$^+$ & $3.55(-4)$ & $4.88(-21)$ & $-0.42$ & 1.7 \\
Si$^+$& $3.35(-5)$ & $3.42(-22)$ & $-0.12$ & 1.4 \\
S$^+$ & $1.86(-5)$ & $1.51(-25)$ & $0$ & 4.7\\
H$^+$ & $1.00(-6)$ & $4.76(-24)$ & $1.68$ & 0.6 \\
\hline 
\end{tabular}
\end{table}

If H$_2$ is predominantly formed through H$^-$ and destroyed by photodissociation at a rate $k_{photo}$, the steady-state H$_2$ abundance is
\begin{equation}
x(\rm{H}_2) = R^{\rm{H}_2}_{elec}/(\nH k_{photo}).
\end{equation}
For an unsheilded Mathis radiation field rescaled by a factor G$_0$, the steady-state H$_2$ abundance yields\footnote{Assuming an H$_2$ photodissociation rate of $k = 6 \times 10^{-11} G_0$~s$^{-1}$.}
\begin{equation}
x(\text{H}_2) = 5.6 \times 10^{-9} \frac{x_e}{4.8~10^{-4}} \left( \frac{T_{\rm{K}}}{1000 \rm{K}}\right)^{0.9}   \frac{\nH}{G_0} \frac{\eta}{1+\eta},
\label{eq:analytic-H-}
\end{equation}
with $\eta$ given in eq. (\ref{eq:eta-def}).

\paragraph{Ionic catalysis.} H$_2$ formation can also be catalyzed by any ion noted here X$^+$, through the intermediate ion XH$^+$. The process is made of a slow radiative association and a fast ion neutral reaction:
\begin{align}
&\rm{X}^+ + \Hy \rightarrow \rm{X}\Hy^+ + h \nu & k_{1}^X,  \label{eq:rad-asso-XH+}  \\
&X\Hy^+ + \Hy \rightarrow \Hy_2 + \rm{X}^+  & k_{2}^X. \label{eq:rec-XH+} 
\end{align}
This route is the analogous of the former with ions being the catalyst. As for the latter, the efficiency of the catalytic process depends on the survival of X$\Hy^+$. In protostellar winds, the main destruction routes that can compete with reaction (\ref{eq:rec-XH+}) are dissociative recombination that reduces the efficiency of the catalysis by a factor less than 30$\%$, and by photodissociation. Unlike the fragile H$^-$, ionic intermediates X$\Hy^+$ are photodestroyed by UV photons ($\lambda <$ 400~nm). For an unshielded Mathis radiation field, reaction (\ref{eq:rec-XH+}) dominates over photodissociation if
\begin{equation}
    \nH/G_0 > \nH^{crit, X} \equiv \frac{k^{0,X}_{\phi}}{k_2^X}~~,
    \label{eq:eta} 
\end{equation}
where $\nH^{crit, X}$ is given in Table \ref{table:H2-rates} for the main ions, and $k^{0,X}_{\phi}$ is the photodissociation rate of X$\Hy^+$ by an unshielded Mathis radiation field. Below this critical value, photodissociation reduces the efficiency of the considered route. Defining
\begin{equation}
\eta_{X} \equiv \frac{\nH}{G_0  \nH^{crit, X}},
\end{equation}
a fraction 
\begin{equation}
\eta_{X} \equiv \frac{\nH}{G_0  \nH^{crit, X}} 
\end{equation}
of the XH$^+$ formed by radiative association is converted in H$_2$.
As for H$^-$, X$\Hy^+$ ions reach steady-state in very short time scales and yield to a total formation rate of H$_2$ by X$^+$ that can be written as
\begin{align}
    R_{X^+} = \alpha \left(\frac{x(\text{X}^+)}{x_X}\right) \left( \frac{T}{1000 K} \right)^{\beta} \frac{\eta_X}{1+ \eta_X}  \left(\frac{\nH}{1 \rm{cm}^{-3}} \right)^2~~,
    \label{eq:ionic-catalysis}
\end{align}
where $x_X$ is a reference value for the abundance $x(\text{X}^+)$ of the catalyst X$^+$, and $\alpha$, $\beta$ are given in Table \ref{table:H2-rates}. 

\subsection{Comparison between routes}

\paragraph{Dust-free.}

Formation rates of H$_2$ by H$^-$ and XH$^+$ deduced from parameters and reported in Table \ref{table:H2-rates} allow to directly compare the efficiency of the ionic and electron catalytic routes. At 1000~K, and for the ionic and electron abundances reported in Table \ref{table:H2-rates}, the order in which species dominate is e$^-$, C$^+$, S$^+$, Si$^+$, H$^+$. Note that depending on the abundance of the catalyst and on the temperature, this order can vary. For example, formation by H$^+$ takes over formation by C$^+$ for $x(H^+) = 10^{-5}$ and $T_{\rm{K}} \ge 10000~\text{K}$. However, in dust-free winds, the main physical parameter that changes this order is the photodestruction of the fragile intermediates anion H$^-$ by the visible field. Indeed, as shown by eq. (\ref{eq:photo-det-condition}), the photodetachment of H$^-$ reduces the efficiency of this route for
\begin{align}
\nH/W < \nH^{crit} = 4.6 \times 10^{14} \rm{cm}^{-3}. 
\label{eq:boundary-2}
\end{align}
This define the boundary \ding{173} in Fig. \ref{fig:H2-summary}. Though, at this critical $\nH/W$, formation though H$^-$ still dominates over CH$^+$ route. Assuming that CH$^+$ is not photodestroyed ($n_{\rm{H}}/G_0>1.7$\dens), formation by C$^+$ takes over formation by e$^-$ if
\begin{align}
R_{\text{C}^+} \ge R_{elec}.
\end{align}
Combining eq. (\ref{eq:ionic-catalysis}) and (\ref{eq:formation-elec-cata}), this condition yields
\begin{align}
n_{\text{H}}/W  \le 6.7\times 10^{12} \left(\frac{x(\text{C}^+)}{3.6~10^{-4}}\right) \left(\frac{x_e}{4.8~10^{-4}}\right)^{-1} \left( \frac{T_{\rm{K}}}{1000 \text{K}} \right)^{-1.32} \text{cm}^{-3}.
\label{eq:boundary-3}
\end{align}
This equation gives the boundary \ding{174} of Fig. \ref{fig:H2-summary}.

H$_2$ can also be formed by three-body reactions with a rate
\begin{align}
R_{3B} = k^{3B} n_{\rm{H}}^3.
\label{eq:3bodyrate}
\end{align}
Assuming that electron catalysis is optimal and using eq. (\ref{eq:3bodyrate}) and (\ref{eq:formation-elec-cata}), we find that three-body reaction takes over formation through H$^-$ if
\begin{equation}
n_{\text{H}} \ge 1.9 \times 10^{13}~ \left( \frac{x_e}{4.8~10^{-4}}  \right) \left(\frac{T_{\rm{K}}}{1000~\text{K}}\right)^{1.24}~\text{cm}^{-3}.
\label{eq:boundary-1}
\end{equation}
This defines the boundary \ding{172}.

\paragraph{Dusty.}
The inclusion of dust does not change the efficiency of gas-phase formation routes but add a new formation route with a rate of
\begin{equation}
R_{dust} = 3.6 \times 10^{-17} \frac{Q}{Q_{\text{ref}}} \frac{S(T_{\text{K}})}{S(1000~\text{K})} \sqrt{\frac{T_{\rm{K}}}{1000~\text{K}}} n_{\rm{H}}^2 ~\rm{cm}^{-3} s^{-1},
\label{eq:dustrate}
\end{equation}
where the formation rate of \cite{1979ApJS...41..555H} assuming a single grain size distribution of radius $a_g = \sqrt{<r_c^2>} = 20$~nm and a grain temperature of $15$~K is adopted. $S(T_{\text{K}})$ is the sticking coefficient of H on grain:
\begin{equation}
S(T_{\rm{K}}) =  \frac{1}{1+0.4\left(\frac{T_{\rm{K}}}{100~\rm{K}} + \frac{T_{\rm{gr}}}{100~\rm{K}}\right)^{0.5} + 0.2 \frac{T_{\rm{K}}}{100~\rm{K}}  + 0.08 \left( \frac{T_{\rm{K}}}{100~\rm{K}} \right)^2 },
\label{eq:stickingproba}
\end{equation}
where $T_{\rm{gr}}$ is the grain temperature. \REV{The sticking coefficient does not depend on $T_{\rm{gr}}$ as long as $T_{\rm{gr}} \lesssim T_K$ and we adopt $T_{\rm{gr}} = 15$~K for simplicity.}

When gas-phase formation is dominated by H$^-$, the critical dust fraction ratio above which formation on dust dominate is obtained by combining eq. (\ref{eq:dustrate}) and (\ref{eq:formation-elec-cata}):

\begin{align}
Q/Q_{\text{ref}}  \ge 9.3 \times 10^{-3} \left( \frac{x_e}{4.8~10^{-4}} \right) & \left( \frac{T_{\text{K}}}{1000~K} \right)^{0.4} \left(\frac{S(T_{\text{K}})}{S(1000~\text{K})} \right)^{-1} \frac{\eta}{1+\eta}
\label{eq:boundary-56}
\end{align}
where $\eta$ is defined in eq. (\ref{eq:eta-def}).
Boundary \ding{177} corresponds to the case $\eta \gg 1$, and boundary \ding{176} corresponds to the case $\eta\ll 1$.
When gas-phase formation route is dominated by CH$^+$, formation on dust takes over gas-phase formation for

\begin{align}
Q/Q_{\text{ref}}  \ge 1.4 \times 10^{-4} \left( \frac{x(\rm{C}^+)}{3.6~10^{-4}} \right) & \left( \frac{T_{\text{K}}}{1000~K} \right)^{-0.92} \left(\frac{S(T_{\text{K}})}{S(1000~\text{K})} \right)^{-1}. 
\label{eq:boundary-4}
\end{align}
This relation defines the boundary \ding{175}.

\end{appendix}

\end{document}